\documentclass[fleqn,usenatbib]{mnras}
\usepackage[T1]{fontenc}
\usepackage{ae,aecompl}
\usepackage{graphicx}
\usepackage{amsmath}
\usepackage{amssymb}
\usepackage{hyperref}
\usepackage{pdflscape}
\usepackage{svg}

\title[MeerKAT L-band beam modelling]{Primary beam effects of radio astronomy antennas -- II. Modelling MeerKAT L-band beams}

\author[K. M. B. Asad et al.]{K. M. B. Asad,$^{1,2,3,4}$\thanks{E-mail: kasad@iub.edu.bd}
J. N. Girard,$^{5}$
M. de Villiers,$^{4}$
T. Ansah-Narh,$^{2}$
K. Iheanetu,$^{2}$
\newauthor
O. Smirnov,$^{2,4}$
M. G. Santos,$^{3,4}$
R. Lehmensiek,$^{6}$
J. Jonas,$^{2,4}$
D. I. L. de Villiers,$^7$
\newauthor
K. Thorat,$^{2,4,8}$
B. Hugo,$^{2,4}$
S. Makhathini$^{2}$,
G. I. G. Jozsa$^{2,4,9}$
and S. K. Sirothia$^{2,4}$
\\
$^1$ARGI, Department of Physical Sciences, Independent University, Bangladesh, Bashundhara RA, Dhaka, Bangladesh\\
$^2$Department of Physics and Electronics, Rhodes University, PO Box 94, Grahamstown, 6140, South Africa\\
$^3$Department of Physics and Astronomy, University of the Western Cape, Bellville, Cape Town, 7535, South Africa\\
$^4$South African Radio Astronomy Observatory, 2 Fir Street, Black River Park, Observatory, Cape Town, 7405, South Africa\\
$^5$AIM, CEA, CNRS, Université Paris-Saclay, Université de Paris, F-91191 Gif-sur-Yvette, France\\
$^6$ EMSS Antennas, Stellenbosch, South Africa\\
$^7$ Department of Electrical and Electronic Engineering, Stellenbosch University, Stellenbosch, South Africa\\
$^8$ Department of Physics, University of Pretoria, Hatfield, Pretoria, 0028, South Africa\\
$^9$ Argelander-Institut f{\"u}r Astronomie, Auf dem Hu{\"u}gel 71, D-53121 Bonn, Germany
}

\date{Accepted XXX. Received YYY; in original form ZZZ}

\pubyear{2020}

\begin{document}
\label{firstpage}
\pagerange{\pageref{firstpage}--\pageref{lastpage}}
\maketitle

\begin{abstract}
After a decade of design and construction, South Africa's SKA-MID precursor MeerKAT has begun its science operations.
To make full use of the widefield capability of the array, it is imperative that we have an accurate model of the primary beam of its antennas.
We have taken available L-band full-polarization `astro-holographic' observations of three antennas and a generic electromagnetic simulation and created sparse representations of the beams using principal components and Zernike polynomials.
The spectral behaviour of the spatial coefficients has been modelled using discrete cosine transform.
We have provided the Zernike-based model over a diameter of 10 degrees averaged over the beams of three antennas in an associated software tool (EIDOS) that can be useful in direction-dependent calibration and imaging.
The model is more accurate for the diagonal elements of the beam Jones matrix and at lower frequencies.
As we get more accurate beam measurements and simulations in the future, especially for the cross-polarization patterns, our pipeline can be used to create more accurate sparse representations of MeerKAT beams.
\end{abstract}


\section{Introduction}
Observational radio astronomy is continuing its growth through the construction of new generations of radio telescopes such as LOFAR \citep{LOFAR2013}, ASKAP \citep{ASKAP2016}, MeerKAT \citep{Jonas2016} and the upcoming SKA \citep{SKA2015}.
The new and upcoming telescopes can offer exquisite sensitivity and resolution and the ability to image large fractions of the sky very quickly which makes them ideal for exploring new science that relies on the detection of extremely weak astrophysical signals.
However, to make full use of these enhanced capabilities, observers must improve their calibration techniques and the models of instrumental effects.
The first and second generations of calibration strategies \citep[defined in][]{Smirnov2011a} can no longer do justice to the new interferometers.

The key science goals of these newcomers will often demand an implementation of more advanced calibration strategies where both direction independent (e. g., receiver electronics) and dependent (e. g., primary beam and ionosphere) effects are taken into account under an unified mathematical formalism such as the radio interferometer measurement equation \citep[RIME:][]{Hamaker1996,Smirnov2011a} which relates the visibilities observed ($\mathbf{V}_{pq}$) by a baseline $pq$ (formed by the antennas $p$ and $q$) to the true sky represented by the `brightness matrix' $\mathbf{B}$.
It is given by the equation \citep[following][eqn. 18]{Smirnov2011a}
\begin{equation}
\mathbf{V}_{pq} = \mathbf{G}_p \left( \iint\limits_{lm} \mathbf{E}_p \mathbf{B} K_{pq} \mathbf{E}_q^H \frac{dl dm}{n} \right) \mathbf{G}_q^H
\end{equation}
where $K_{pq}=e^{-2\pi i (u_{pq}l +v_{pq}m)+w_{pq}(n-1)]}$, $u,v,w$ are the coordinates of the baseline in a reference frame oriented toward the observing direction, $\mathbf{G}$ and $\mathbf{E}$ are their direction independent (DIE) and dependent systematic effects (DDE), respectively, $l,m$ are the `direction cosines' toward the sky and $n=\sqrt{1-l^2-m^2}$. In this paper, $\mathbf{E}$ will be used to denote only the primary beam (hereafter, only \textit{beam}, which should not be confused with the \textit{point spread function} of an array), neglecting other DDEs, such as the ionospheric, tropospheric and Faraday rotation effects.

The worrying assumption of traditional self-calibration, i. e. second generation calibration, is that DDEs are trivial, which implies each visibility is a measurement of the sky `coherency function' corrupted by some \textit{multiplicative} gains.
In such a scenario, an interferometer array would measure the Fourier transform of one \textit{common} sky.
This assumption holds only if the DDEs are identical across all antennas and constant in time, which they are not.
In the presence of the actual non-trivial DDEs, the observed visibilities would be a \textit{convolution} of the `sky coherency' and the DDEs \citep{Smirnov2011b}.
In our case, the beam convolves the ideal visibilities with a different kernel per antenna and per time and frequency step and, hence, we get a different $uv$-plane for every baseline, time and frequency sample.
Correcting for the beam effects is not trivial and still under much experimentation \citep[for an example, see][]{Bhatnagar2013,Tasse2018}, but a clear pre-requisite step towards this complex task would be making an accurate beam model.

A simple \textit{approximate} model can be created using, e. g., Gaussian or Jinc functions that can capture the basic features of the beam.
However, such models cannot account for asymmetries in the shapes of the main and side lobes and the position and level of far-out sidelobes.
Capturing such spectral effects will be crucial to properly characterise weak astrophysical signals.
For instance, the foreground cleaning techniques in the intensity mapping of neutral Hydrogen (HI) during the epoch of reionization \citep{Mellema2013} and also at lower redshifts \citep{Santos2015} rely on the spectral smoothness of the diffuse Galactic foregrounds, which can be destroyed by such frequency effects if the beam is not properly modelled and corrected for.

More \textit{realistic} models of the beam, in angular space, time and frequency, can be either \textit{phenomenological}, based on the appearance of the far-field radiation pattern of an observing antenna, or \textit{physical}, based on the actual engineering physics of the antenna \citep[described briefly in][section 3.1]{Jagan2018}.
One of the most widely used \textit{phenomenological} methods is `astro-holography' (AH), where a holographic measurement of the far-field pattern of an antenna \textit{scanning} an astronomical source is taken, by cross-correlating its measured voltages with that of another antenna \textit{tracking} the same source \citep[for more information see][]{Morris1988,Cotton1994,Harp2011,Perley2016}.
Measurement of the beam directly through this method is usually called `beam holography' and using the measured beam to model the figure of the reflecting surface of an antenna, through the Fourier relationship between the far-field pattern and the aperture illumination function, is usually called `dish holography'.
The first use of AH in radio astronomy was aimed at the latter \citep{Scott1977}.
However, the basis of most \textit{physical} approaches is the electromagnetic (EM) simulation of an antenna.
EM simulations can predict the beam over a large field of view relatively easily, but it is computationally very expensive to produce EM models for each antenna of an array and for every frequency channel.
In addition, it is more difficult to account for the temporal and environmental behaviour of the beam in EM simulations.
On the other hand, beam models created from AH observations are more accurate and easier to obtain for multiple frequency channels and time samples, but they are usually restricted to smaller fields of view and angular resolution.
Therefore, information from both AH observation and EM simulation can lead us to a better representation of the beam.

This paper is the second in a series of papers dealing with the modelling and effects of the primary beams of radio astronomy antennas taking into account both the \textit{physical} and \textit{phenomenological} approaches.
The first paper presented the modelling of the Karl G. Jansky Very Large Array (VLA) beam from AH and compared these models to those created from EM simulations and physical considerations \citep[][hereafter `Paper I']{Iheanetu2019}.
It presented two different techniques of AH beam modelling---`data-driven' modelling using Principal Component Analysis (PCA) and `basis-driven' modelling using Zernike polynomials (ZP).
In this paper, we will present different approaches of modelling available AH measurements and EM simulations of \textit{MeerKAT}, an SKA-MID precursor array located in South Africa.
Here, we call the data-driven approach `characteristic' and the basis-driven approach `analytic.'
Besides the PCA and ZP approaches, we will also demonstrate the use of spherical harmonics (SH).
The characteristic and analytic basis models created from the AH observations will be compared with each other and also with the EM simulations.
We will also present spectral modelling of the spatial coefficients using Discrete Cosine Transform (DCT).

\begin{figure*}
	\centering
	\includegraphics[width=0.33\linewidth]{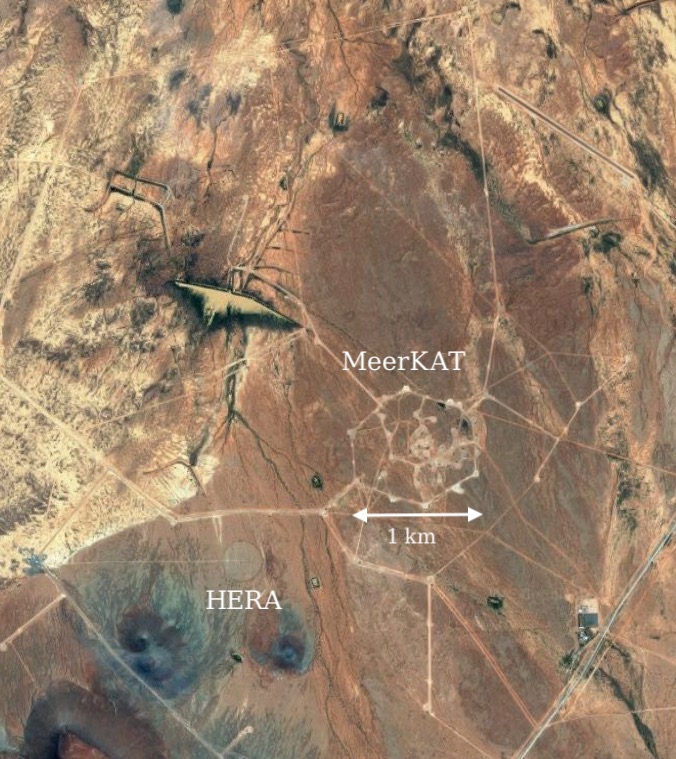}
	\includegraphics[width=0.66\linewidth]{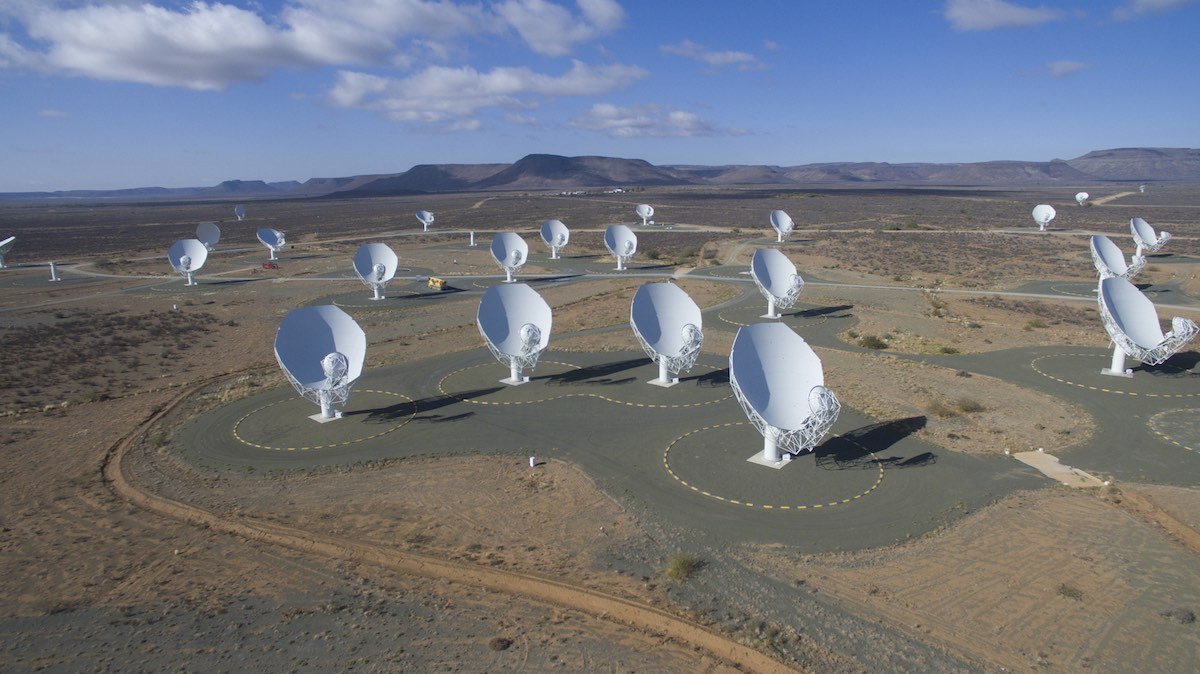}
	\caption{\textit{Left:} A satellite view of the MeerKAT receptor pads before the installation of the receptors, with the  Hydrogen Epoch of Reionization Array (HERA) in the neighbourhood, captured from Google Earth. \textit{Right:} A photograph of some of the MeerKAT receptors with the flat-topped hills (Karoo Kopies) capped by dolerite sills, a reminder of the Gondwanan past, in the background (courtesy SARAO).}
	\label{f:meerkat}
\end{figure*}

\begin{figure*}
	\centering
	\includegraphics[width=0.9\linewidth]{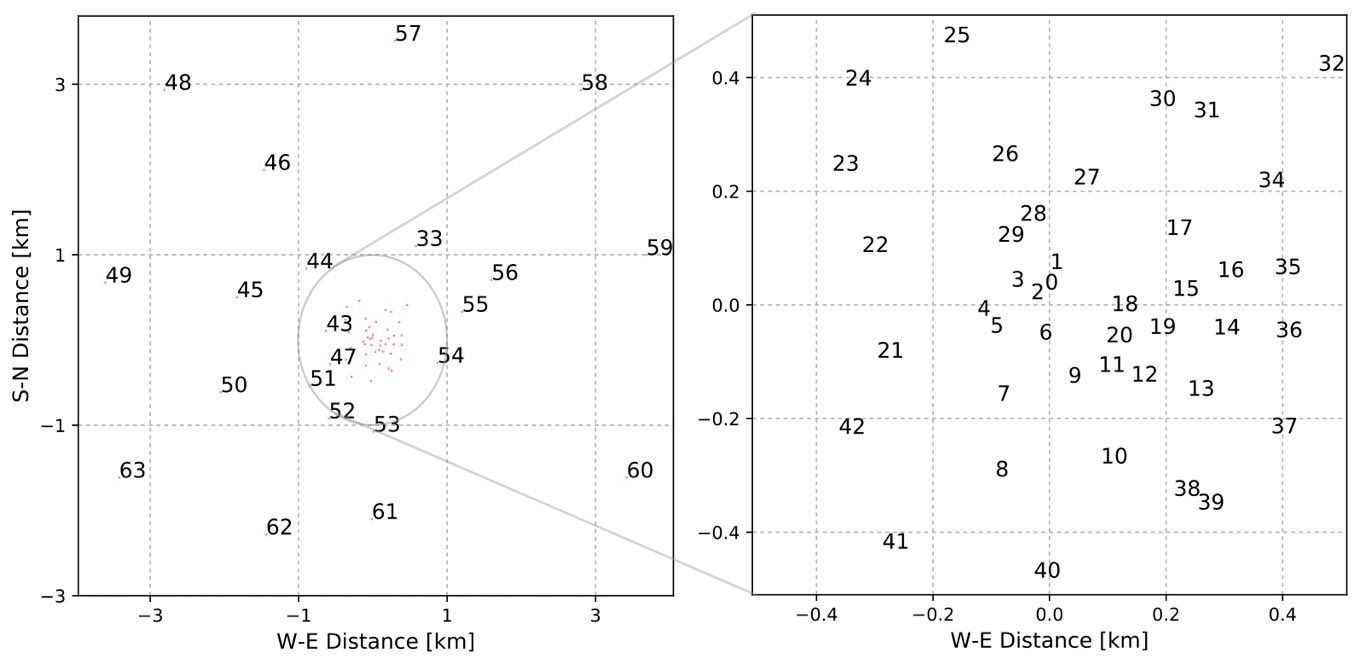}
	\caption{The distribution of the 64 antennas of MeerKAT, each identified by an integer ranging from 0 to 63. Note that the actual names of the antennas are given as M000, M001, M002 and so on. \textit{Left:} The distribution outside the 1-km-core. \textit{Right:} The distribution inside the 1-km-core. The core is loosely delimited by the hexagonal boundary visible in the satellite image of Fig. \ref{f:meerkat} (left). The West-East and South-North distances are shown relative to the arbitrary centre located at $-30^\circ 42' 47.41''$ South, $21^\circ 26' 38.00''$ East.}
	\label{f:meerkat-dist}
\end{figure*}

A simpler and analytic primary beam model of MeerKAT is presented by
\citet{Mauch2020} based on cosine-tapered
field illumination (equation 3 in the paper).\footnote{Available at \url{https://pypi.org/project/katbeam}.}
They have compared this model with an azimuthal average of the AH beam of M. de Villiers (in preparation) and found a good match out to $2.5$ degrees from the phase centre.
We have used the same AH beams and created a more elaborate asymmetric sparse model based upon it.

Section 2 gives a general introduction to MeerKAT and describes the AH observations and EM simulations used in this work.
Section 3 describes the general formalism of the characteristic and analytic approaches of modelling the spatial shape of the beam.
Here, we also discuss spectral modelling of the spatial coefficients and compare the different approaches.
Finally, we end with the main conclusions of the paper and some remarks about our future work in Section 4.
The Zernike-based model presented here is available through the openly accessible tool {\tt EIDOS}.\footnote{\url{https://github.com/ratt-ru/eidos}}

\section{Beam measurement and simulation}
MeerKAT is located in the Upper Karoo region of South Africa.
It has 64 interlinked receptors among which, 48 are located in a core region of 1 km in diameter centred at $-30^\circ 42' 47.41''$ South, $21^\circ 26' 38.00''$ East.
The other 16 are located outside the core giving a maximum baseline of 8 km.
Fig. \ref{f:meerkat} (left) shows a satellite picture of the MeerKAT location and Fig. \ref{f:meerkat} (right) shows a recent photograph of some of its antennas.
The left and right panels of Fig. \ref{f:meerkat-dist} show the distribution of the receptors outside and inside the core, respectively.
We refer the readers to \citet{Jonas2016} and \citet{Camilo2018} for more information about MeerKAT.

Three receivers of MeerKAT are expected to cover three different bands of the radio spectrum, namely the UHF (580--1015 MHz), L (900--1670 MHz) and S (1750--3500 MHz) band.
We will only focus on L-band because substantial AH observations have been carried out at these frequencies.
Beams of all the 64 antennas have been measured at L-band, but we have the observations of only three antennas available for this work.
On the other hand, the EM-simulated beams of MeerKAT have been created from the physical properties of a generic MeerKAT antenna and, hence, it is \textit{assumed} to be same for all antennas.
Because AH and EM beams rely on completely different methods and pipeline, one can be used to check the sanity of another.
In the following subsections, we describe the two beams and use one to check the accuracy of the other.
Antenna-to-antenna variation of the beam for the three antennas, yet another check of the sanity of the beam, is also discussed.

\begin{table}
	\centering
	\caption{Information about the holographic observation analysed in this paper. The data was averaged in frequency to get a resolution of 0.8 MHz and only three among the 18 available scanning antennas were used.}
	\label{t:obs1}
	\begin{tabular}{ll} 
		\hline
Experiment ID & 20181206-0010 \\
Target object & 3C 273 \\
Target RA (J2000) & $12^h 29^m 06.70^s$ \\
Target DEC (J2000) & $+02^\circ 03' 08.6''$ \\
Start time & 6 December, 2018, 10:29:01 SAST \\
Duration & 0.5 hours \\
Time resolution & 1.0 s \\
Scanning antenna & M009, M012, M015 \\
Centre frequency & 1284.0 MHz \\
Bandwidth & 856 MHz \\
Channel width & 0.835937 MHz \\
Number of channels & 1024 \\
Average elevation & $39.82^\circ$ \\
Average azimuth & $303.22^\circ$ \\
		\hline
	\end{tabular}
\end{table}

\subsection{Astro-holographic observation} \label{s:ah}
AH observations are available for all 64 antennas of MeerKAT.
We have taken one such observation in which the nearest bright quasar 3C 273 \citep{Schmidt1963} was scanned using 18 antennas and simultaneously tracked using another 36 antennas for 0.5 hours.
The relevant observational parameters are given in Table \ref{t:obs1}.
It is classified as an `astro'-holographic observation because it was targeted at an astronomical object;
for an example of a holographic observation of the beam of a MeerKAT antenna using a satellite beacon instead of an astronomical source, see \citet[][figure 6]{Jonas2016}.

After calibrating the observed data, the AH beam was extracted for three of the scanning antennas (M009, M012 and M015) using all the available tracking antennas.
The raw noisy beam measurements are then de-noised via aperture-plane masking.
Aperture-plane masking is performed relying on the Fourier relationship between the primary beam and the aperture illumination function (AIF) of an antenna.
The measured beams are Fourier transformed to obtain the AIF, the Fourier modes lying outside the physical aperture plane are masked and, finally, the masked AIF is inverse-Fourier transformed to create a smooth de-noised beam.
We have used the de-noised beam for all analyses in this paper.

For AH observations, the correlator is operated as in normal observing mode (while the antenna tracking strategy is necessarily different), and a set of visibilities is generated.
The calibration process is broadly similar to that of a normal observation, but some important subtleties need to be taken into account (see, e. g., Paper I for an extended discussion).
The data in this work was reduced using a combination of the standard MeerKAT Science Data Processing (SDP) pipeline and the MeerKAT holography tool {\tt KATHOLOG} (M. de Villiers, in prep).

\begin{figure}
	\centering
	\includegraphics[width=\columnwidth]{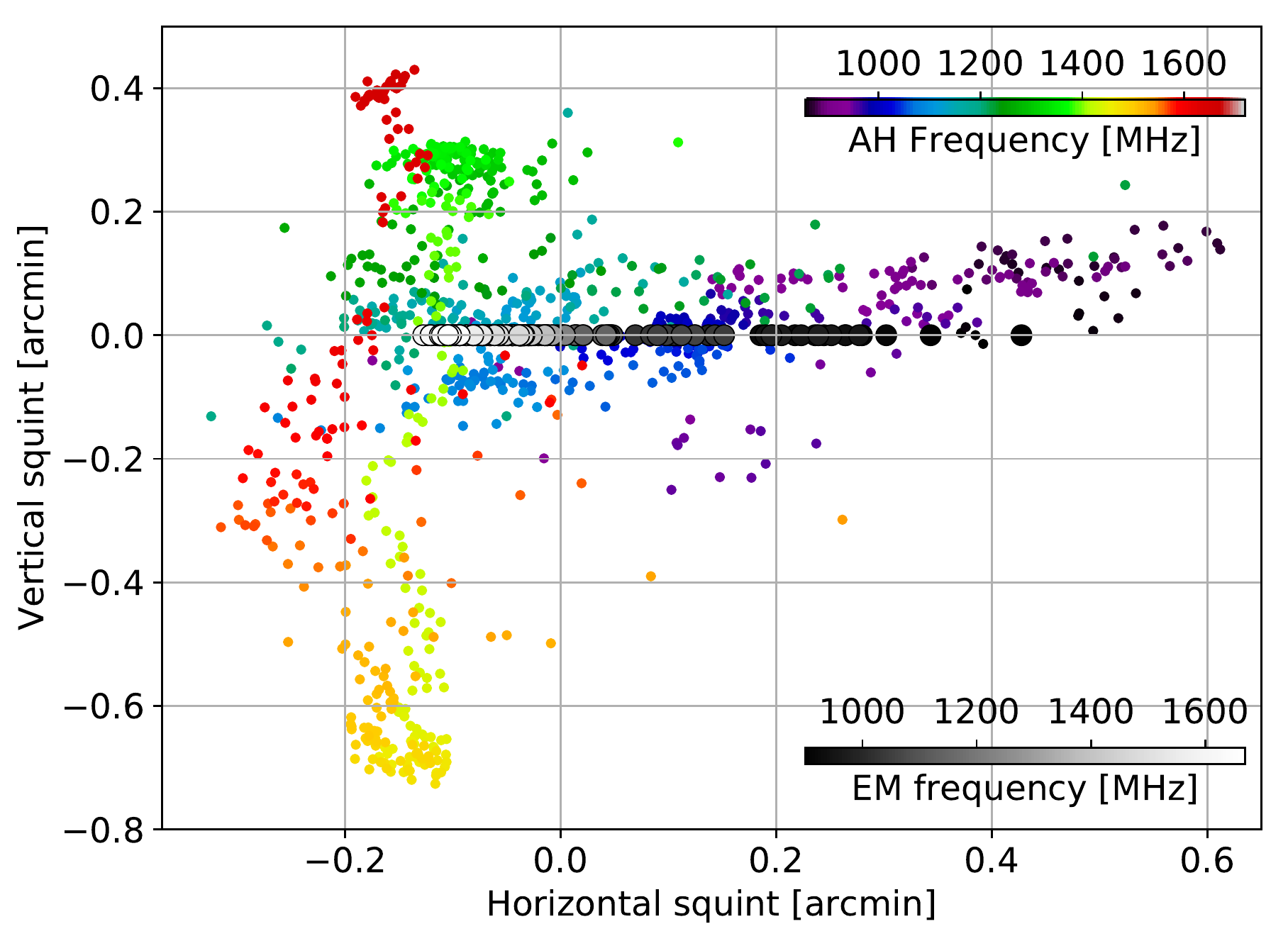}
	\caption{Frequency-dependent squint of the given AH observation (violet to red colours) and EM simulation (shades of grey). The horizontal and vertical squints are calculated from the pointing centre.}
	\label{f:squint}
\end{figure}

\begin{figure}
	\centering
	\includegraphics[width=\columnwidth]{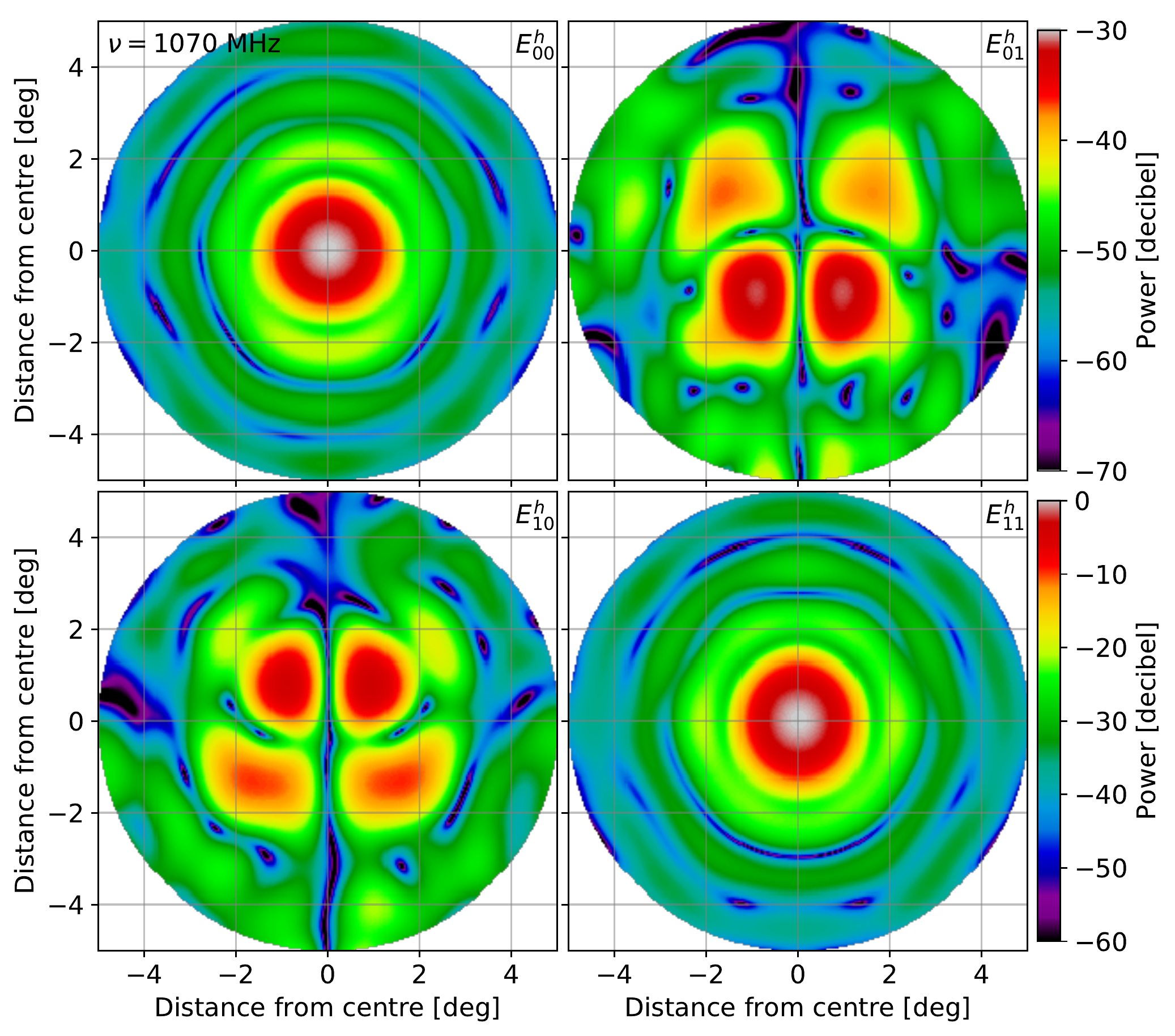}
	\caption{Holographic observation of MeerKAT primary beam over 10 degrees at 1070 MHz averaged over three antennas (M009, M012, M015).
		The observed data has been smoothed by masking Fourier modes in the aperture plane.
		Top and bottom colour-bars are for the off-diagonal and diagonal elements, respectively.
		For details, see Section \ref{s:ah} and Table \ref{t:obs1}.}
	\label{f:ah}
\end{figure}

\begin{figure}
	\centering
	\includegraphics[width=\columnwidth]{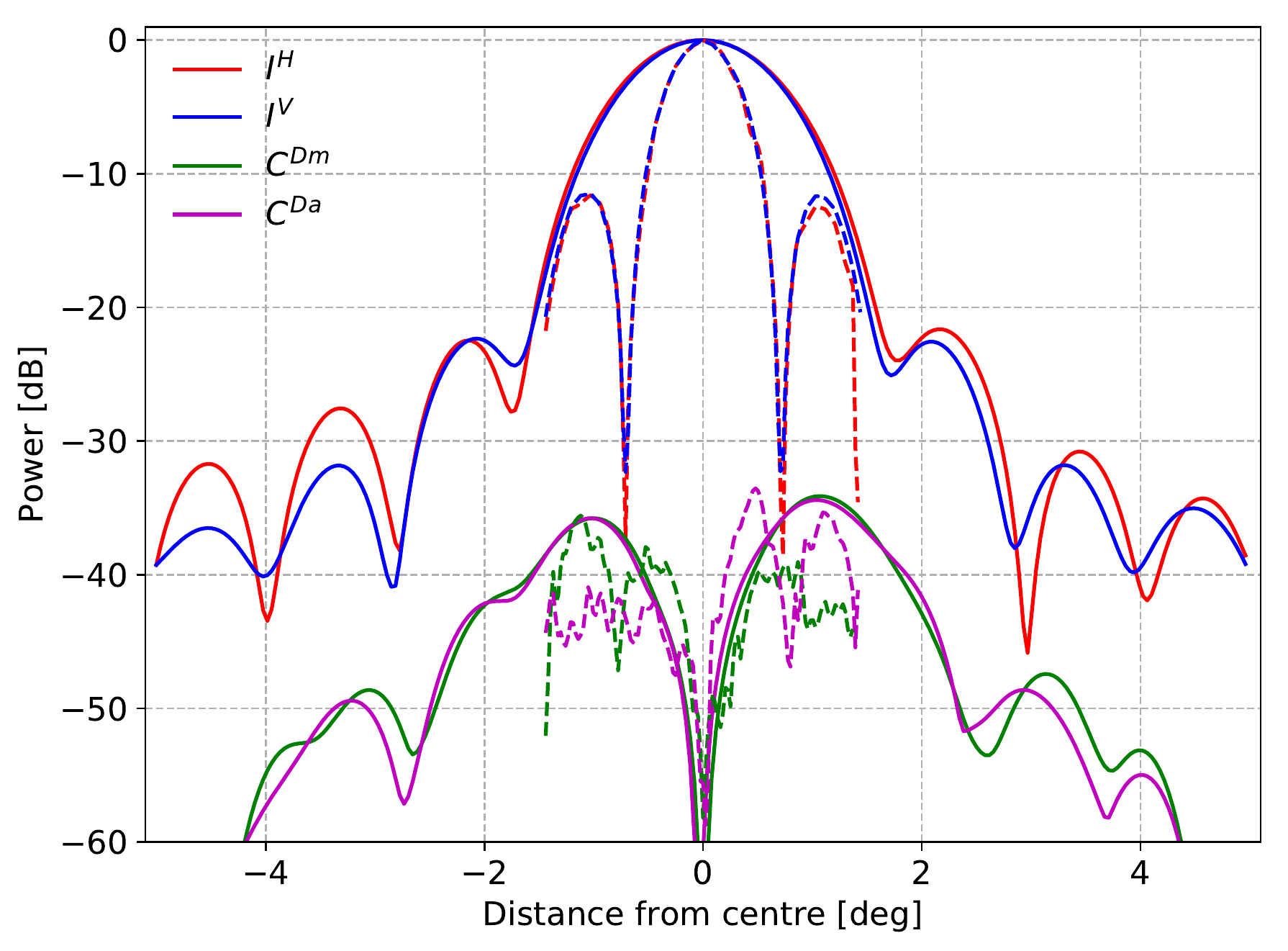}
	\caption{Horizontal ($I^H$) and vertical ($I^V$) cuts through the centre of the Stokes I beam at 1070 MHz and the diagonal cuts ($C^{Dm}$ for the main and $C^{Da}$ for the anti-diagonal) through the centre of the average cross-polarization pattern. The solid lines show the results for MeerKAT and the dashed lines for VLA.}
	\label{f:ah-1d}
\end{figure}

Before producing models from the AH measurements, we shift the beam centres at each frequency channel independently to remove the frequency-dependent offset, i. e. squint, of the beam centre from the pointing centre.
The squint is different for each of the feeds of each antenna and it also varies with frequency.
Therefore, for an accurate comparison between the AH and EM datasets, we remove squint from both of them.
The squints were calculated by fitting 2D elliptical Gaussians on the beam measurements at each frequency.\footnote{The python module {\tt gaussfitter} is used.}
The spectral behaviour of the squints corresponding to the $E_{00}$ element of the AH measurement is shown in Fig. \ref{f:squint}.
Note that the pointing offset averaged over all frequencies is related to the mechanical pointing error of an antenna, not with its optical properties.
Therefore, we have calculated the squint after taking out the average pointing error.
The resulting squint varies horizontally from low to middle frequencies of the L-band, but the variation at higher frequencies is in the vertical direction.
We store the per-frequency per-antenna per-feed squint values and these can be added to the squint-less beam model at a later stage if desired.

The re-centred squint-less beam dataset has the shape ($N^h_\nu\times 2\times 2\times 256 \times 256$) where $N^h_\nu$ is the number of frequency channels, the $2\times 2$ matrices represent the Jones elements, and $256\times 256$ correspond to the total number of pixels.
The beam centre is always at the pixel position $(128,128)$.
We then normalise the beams with respect to the centre by dividing the complex Jones images by the complex Jones matrix formed by the central pixel, i. e. $\mathbf{E} (x,y) = \mathbf{E} (x,y) \cdot \mathbf{E}^{-1}(x=128,y=128)$ where $x,y$ are the pixel coordinates.
Finally, we average the beam measurements of the three antennas to create an antenna-averaged beam because the EM model is created for a generic MeerKAT antenna and, hence, it would be more appropriate to compare this model with an averaged AH beam (the antenna-to-antenna variation is discussed in Section \ref{s:a2a}).

We refer to this re-centred normalised three-antenna-averaged AH beam over a diameter of 10 degrees as $\mathbf{E}^h$.
It is effectively a squint-less differential beam measurement with respect to the centre.
As an example, the squared amplitude of the AH beam at a particular frequency (1070 MHz) is shown in Fig. \ref{f:ah}.
The main lobe and the first three sidelobes in the diagonal elements and a cloverleaf pattern in the off-diagonal elements are visible in the figure.

The sidelobe levels can be seen more clearly in Fig. \ref{f:ah-1d} where a 1D cut through the centre of the Stokes I beam (average of the diagonal Jones elements) is shown.
The asymmetry of the beam is also clear from the difference between the horizontal (red solid line) and vertical (blue solid line) cuts.
The first sidelobe level is more than 20 dB below the maximum and the second sidelobe more than 30 dB below.
Cuts through the main-diagonal (green solid line) and the anti-diagonal (magenta solid line) of the average cross-polarization pattern is also shown.
Cross-polarization power increases with radius and reach a maximum of $\sim -35$ dB in the central region of the cloverleafs.
The corresponding cuts through the VLA beam are shown in the same figure (dashed lines) and we see that the first sidelobe level of VLA beam, at 1070 MHz, is almost an order magnitude higher than that of MeerKAT.
However, their cross-polarization levels are comparable.

\begin{figure}
	\centering
	\includegraphics[width=\columnwidth]{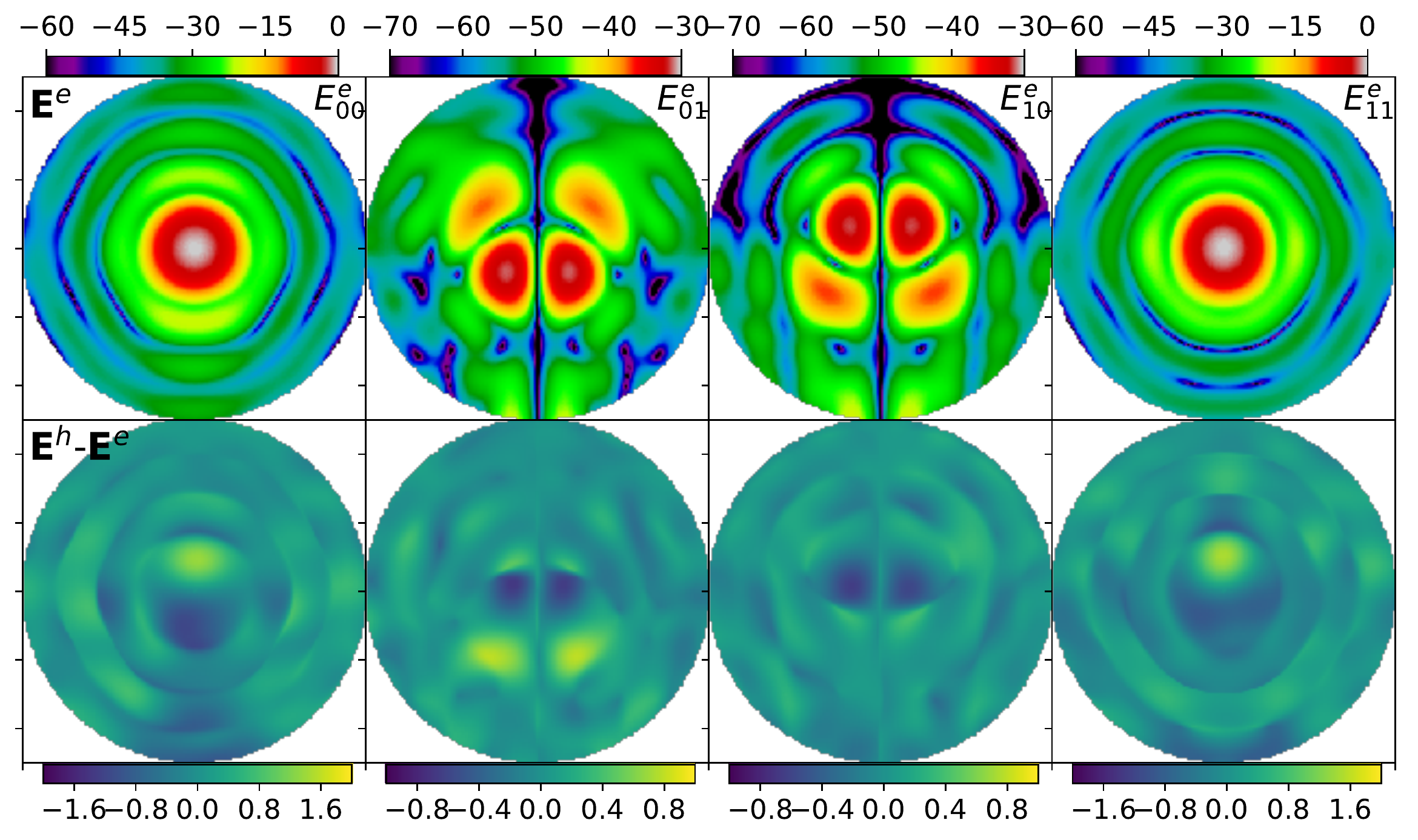}
	
	\includegraphics[width=\columnwidth]{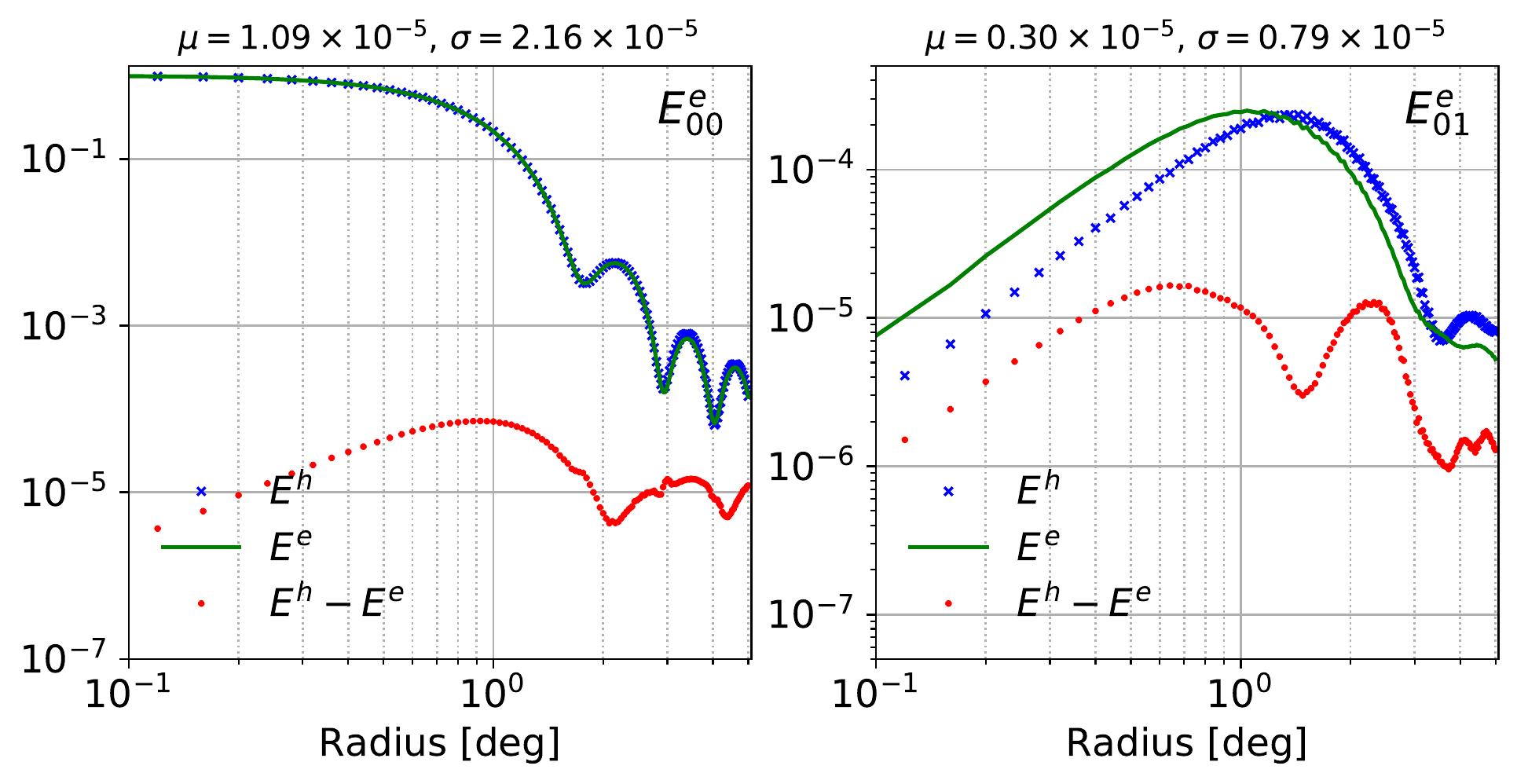}
	\caption{\textit{Top}: Electromagnetic simulation of MeerKAT beam at 1070 MHz over a diameter of 10 degrees (top row) and the residuals after subtracting it from the AH measurement (bottom row). The residuals are multiplied by 100 for better visualisation. \textit{Bottom}: Radial profiles of the $E_{00}$ and $E_{01}$ elements of the AH (blue crosses) and EM (green solid line) datasets and the corresponding residuals (red dots).}
	\label{f:em}
\end{figure}

\begin{figure*}
	\centering
	\includegraphics[width=0.49\linewidth]{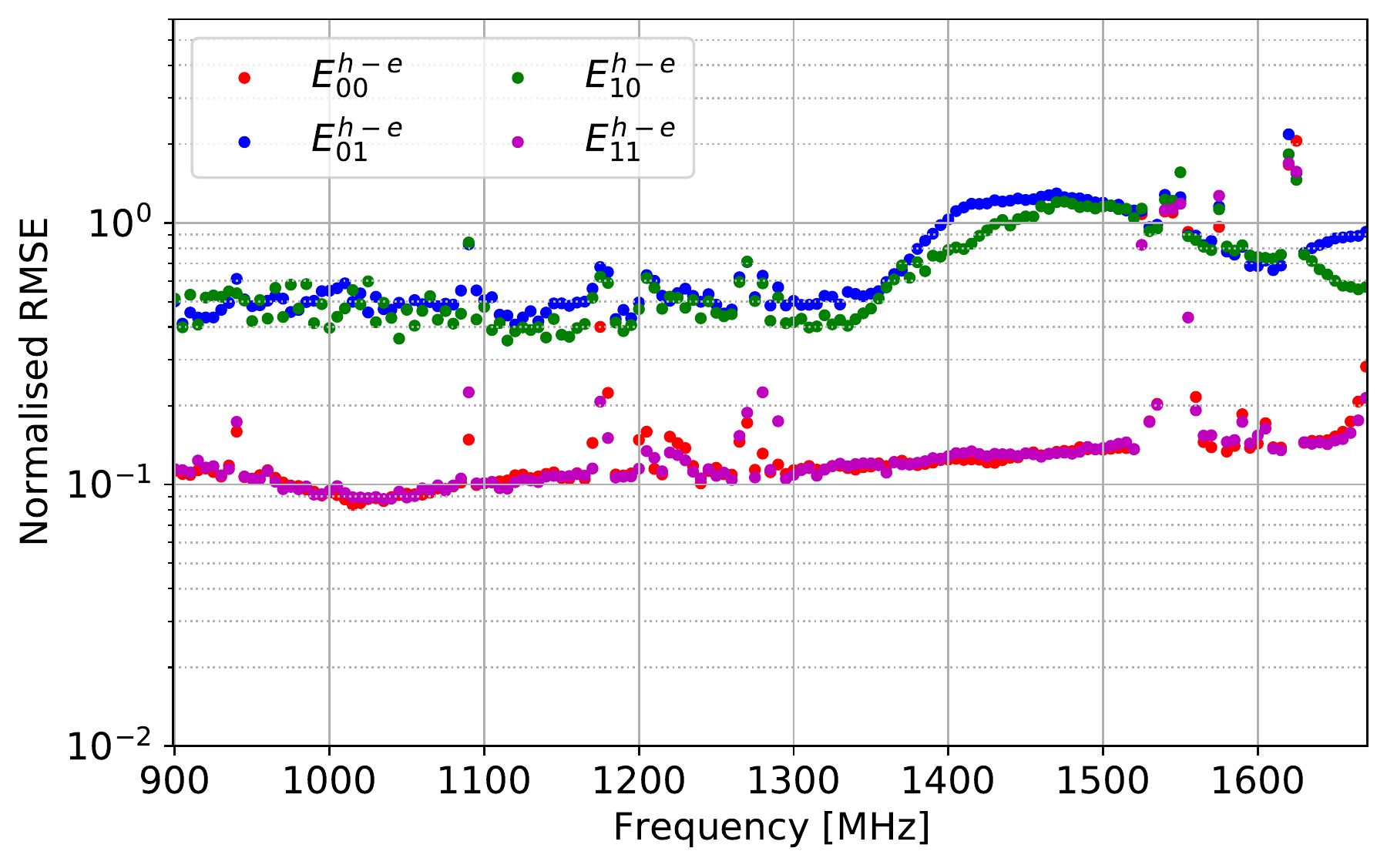}
	\includegraphics[width=0.49\linewidth]{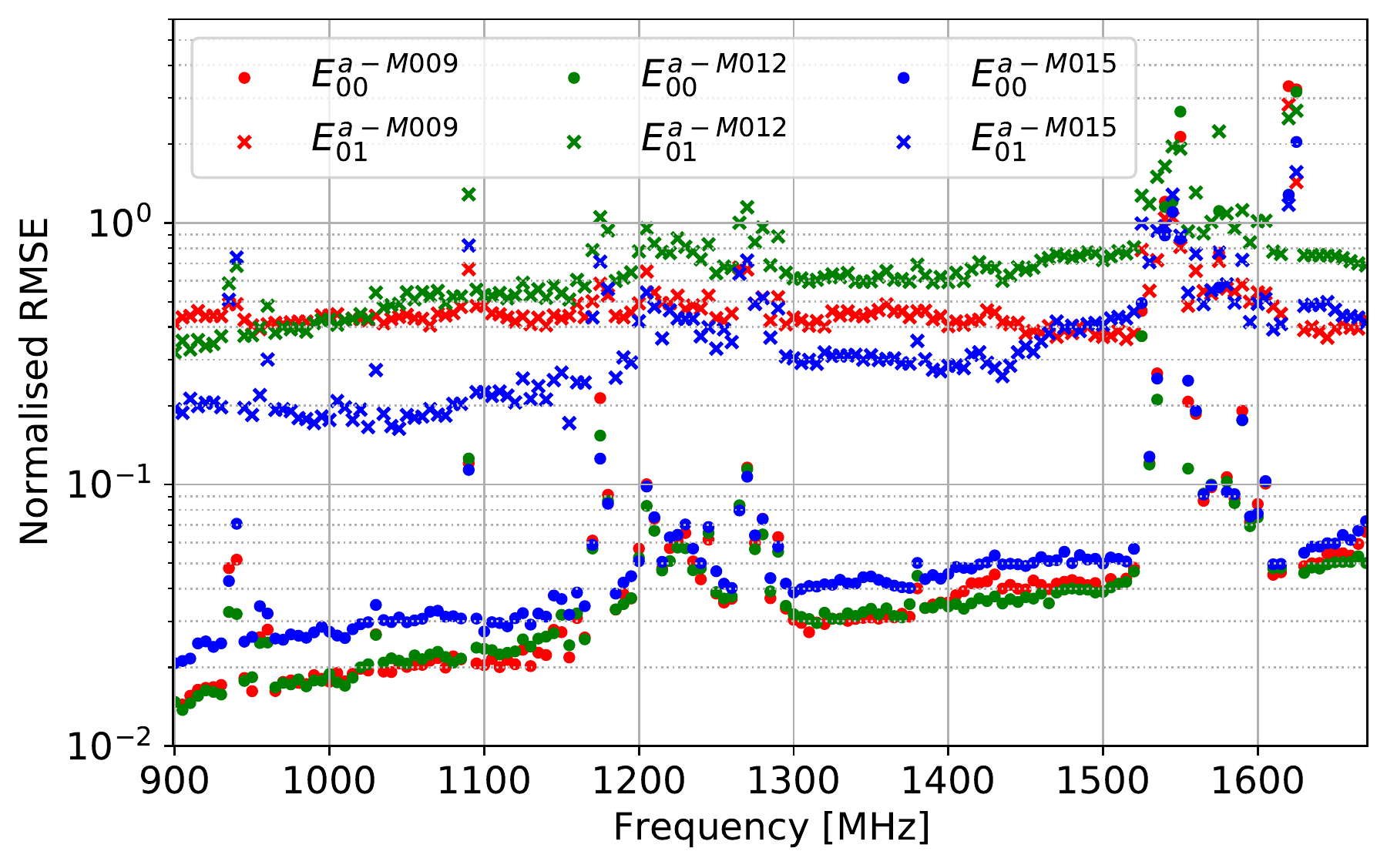}
	\caption{\textit{Left}: Normalised root-mean-square error (NRMSE) of the EM model with respect to the AH observation as a function of frequency averaged over a diameter of 10 degrees. \textit{Right}: NRMSE of the AH beams of three antennas (M009, M012, M015) with respect to the beam averaged over the three antennas ($E^a$). Both the diagonal (dots) and off-diagonal (crosses) elements are shown.}
	\label{f:nrmse-em}
\end{figure*}

\subsection{Electromagnetic simulation}
MeerKAT primary beam has been simulated using the EM simulation software {\tt GRASP}\footnote{\url{https://www.ticra.com/software/grasp}} taking into account the principles of physical optics (PO) and the physical theory of diffraction (PTD).
The {\tt GRASP} simulations compare well with EM simulations performed using the MLFMM (multilevel fast multipole method) technique of {\tt FEKO}.\footnote{\url{https://altairhyperworks.com/product/FEKO}}
The simulations are available for the whole L-band with a spectral resolution of 5 MHz and over the entire hemisphere.
For the purpose of this paper, we restrict ourselves to within a diameter of 10 degrees.
The frequency-dependent squints of the simulated beam are shown in Fig. \ref{f:squint} (shades of grey).
Squint varies smoothly in a horizontal direction as one goes from low to high frequencies and this trend is similar to the AH measurements.
We remove these squints and re-centre the EM models to the same pixel as the centre of the AH measurements and normalise them using the same convention.
The re-centred and normalised EM dataset of shape ($N^e_\nu\times 2\times 2\times 256 \times 256$), where $N^e_\nu$ is the number of channels in the EM simulation, is hereafter referred to as $\mathbf{E}^e$.

The EM model at 1070 MHz is shown in the top row of Fig. \ref{f:em} (top panel, top row) and the residuals after subtracting the model from the corresponding AH measurement is shown in the bottom row.
The residuals have been multiplied by 100 for better visualisation.
There is a low-level dipolar structure in the residuals and further investigation is needed to identify its cause which is not within the scope of this paper.
However, it is certain that the diagonal Jones elements of $\mathbf{E}^h$ and $\mathbf{E}^e$ are much closer to each other than the off-diagonal elements.
The good match in the diagonal and the relatively poor match in the off-diagonal element can be seen more clearly in the bottom panel of Fig. \ref{f:em}, where the radial profiles of the $E_{00}$ and $E_{01}$ elements of the AH and EM datasets and the corresponding residuals are shown.

\begin{figure*}
	\centering
	\includegraphics[width=0.24\linewidth]{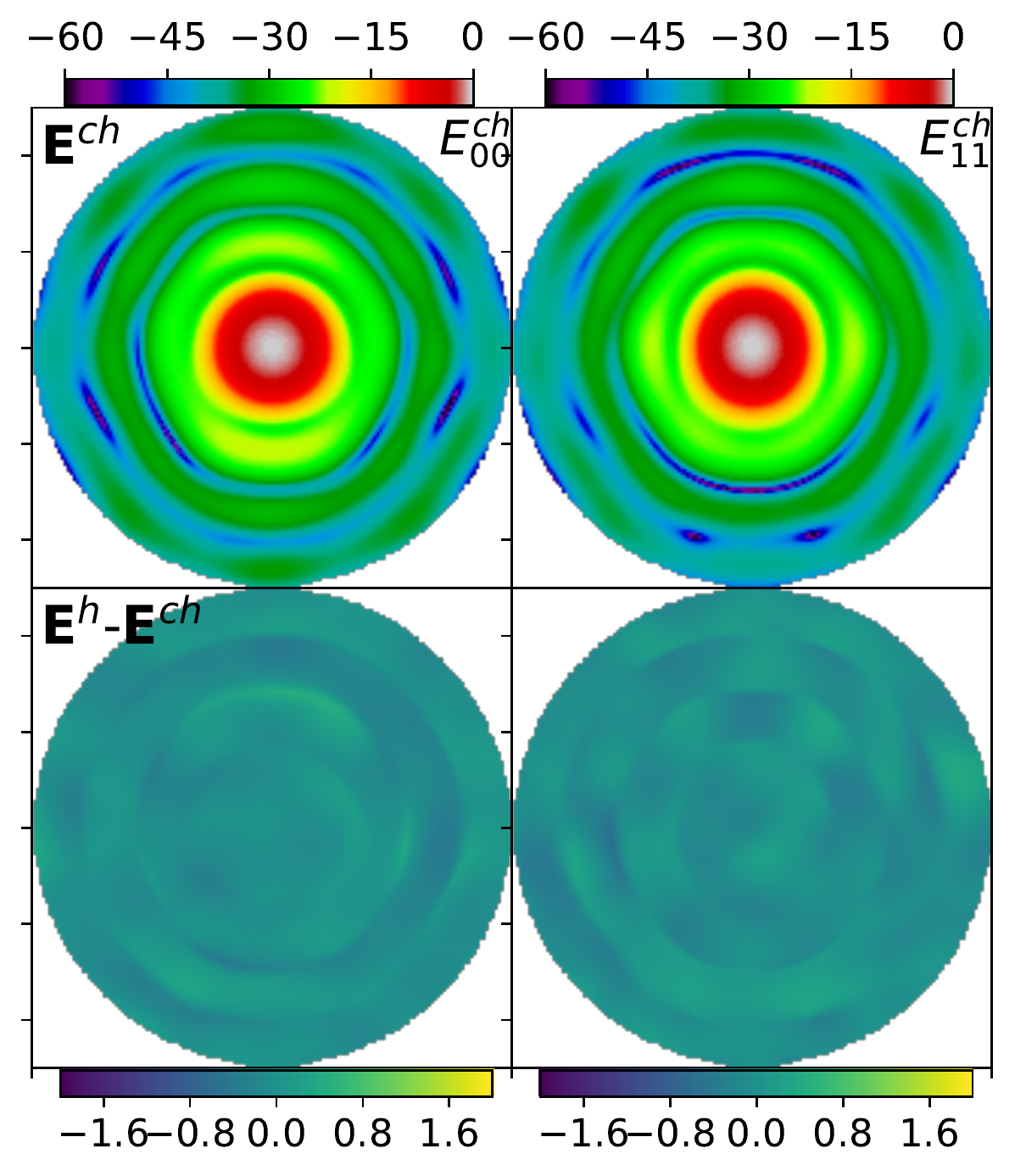}
	\includegraphics[width=0.24\linewidth]{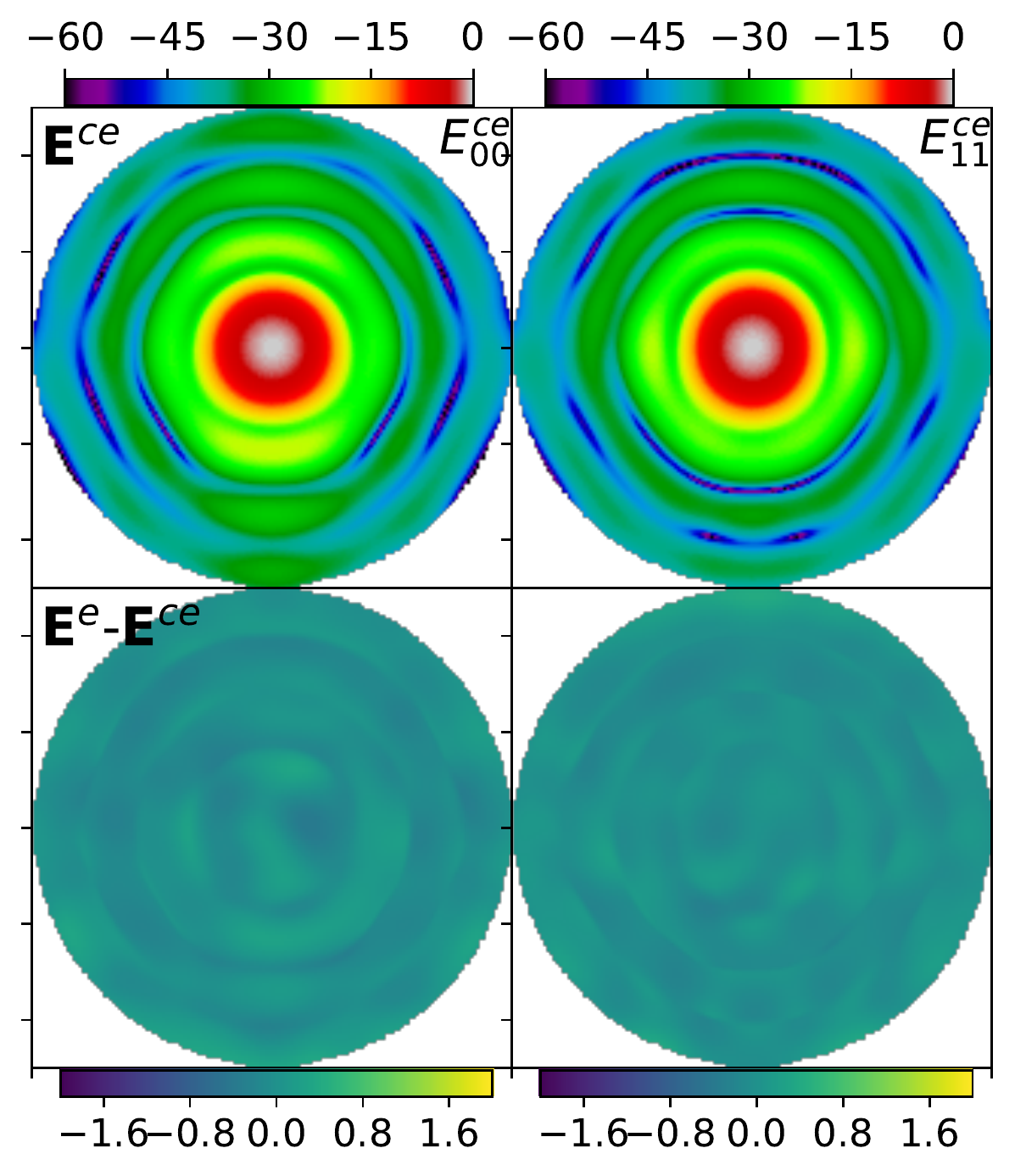}
	\includegraphics[width=0.24\linewidth]{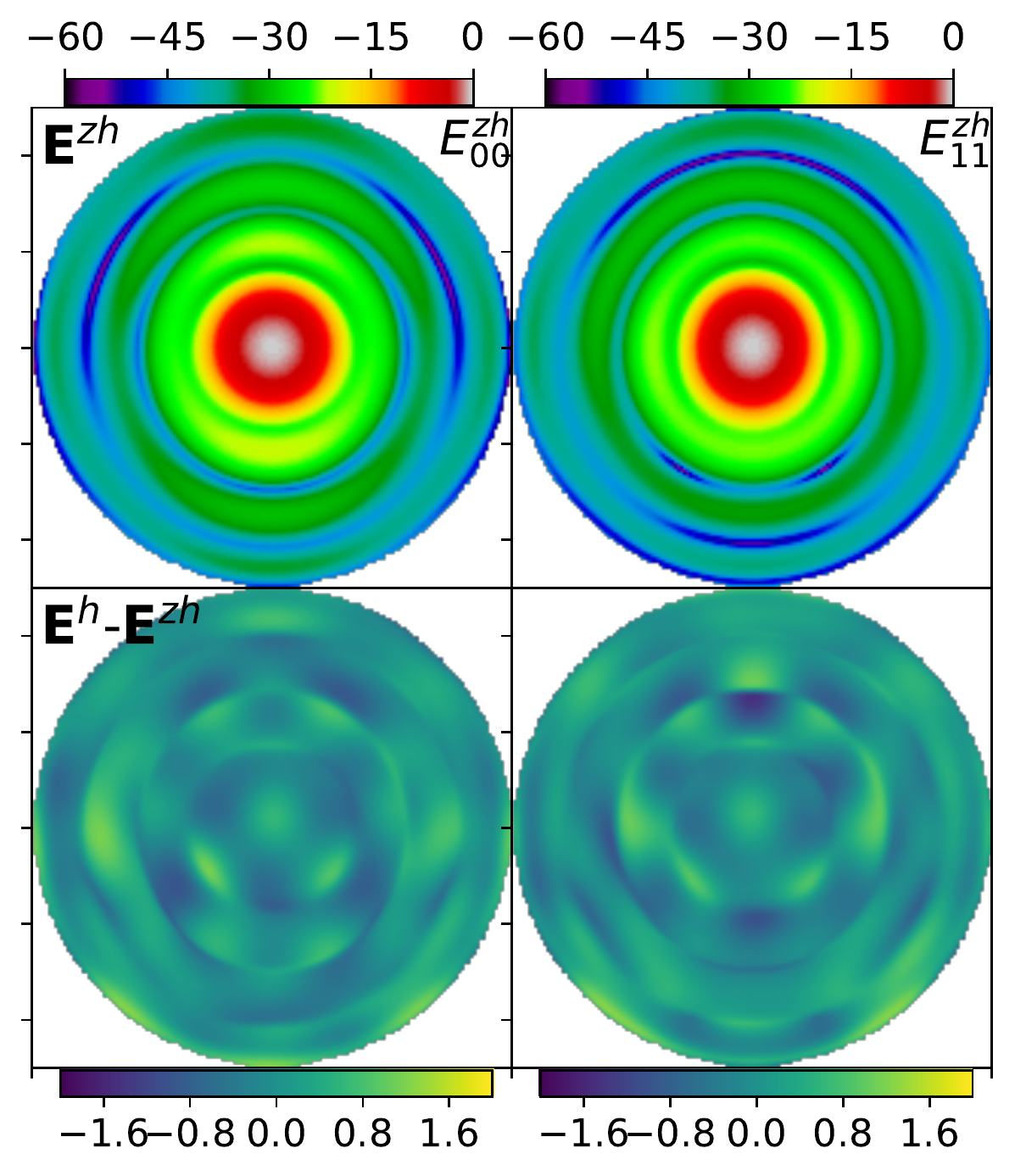}
	\includegraphics[width=0.24\linewidth]{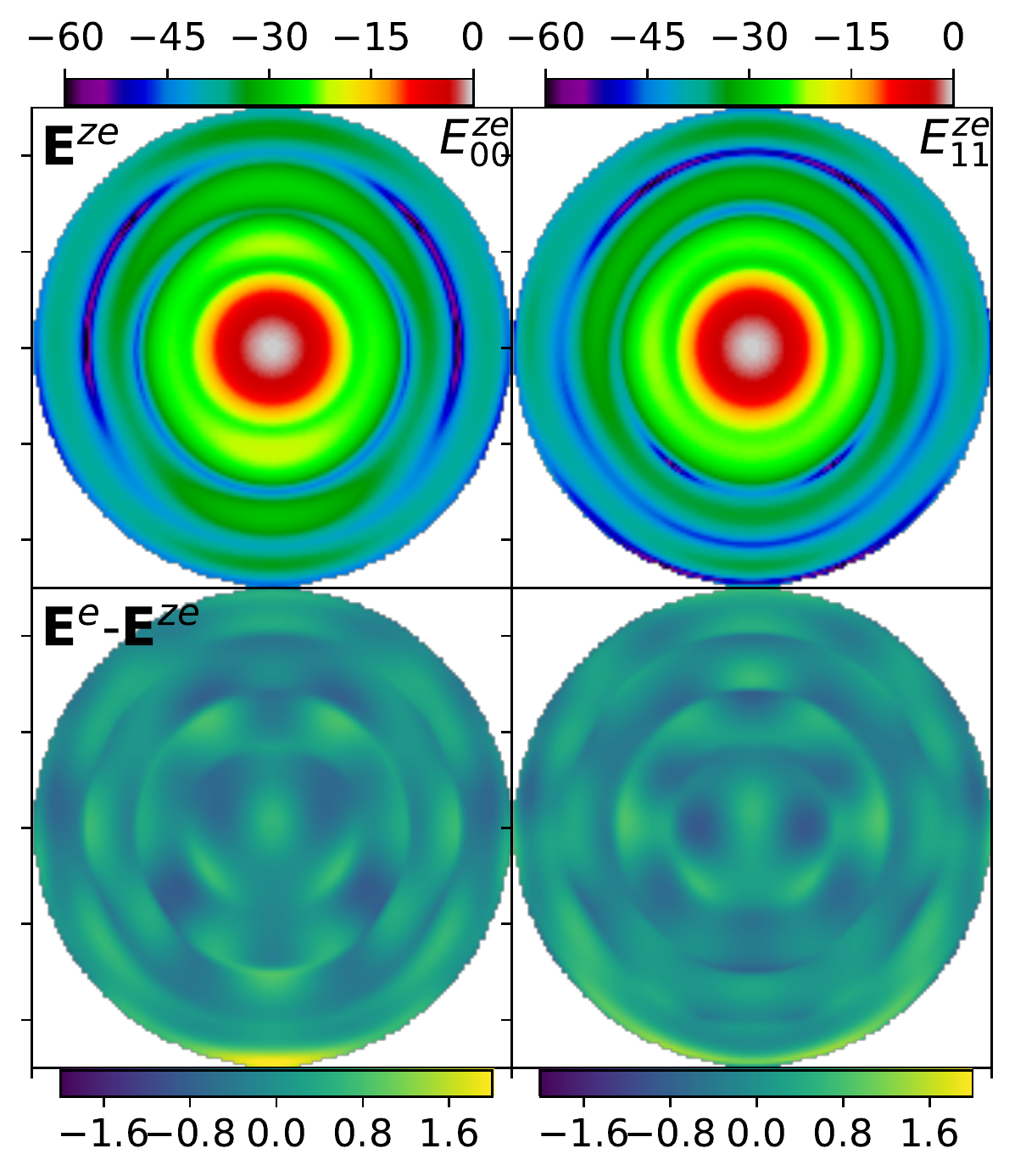}
	
	\includegraphics[width=0.24\linewidth]{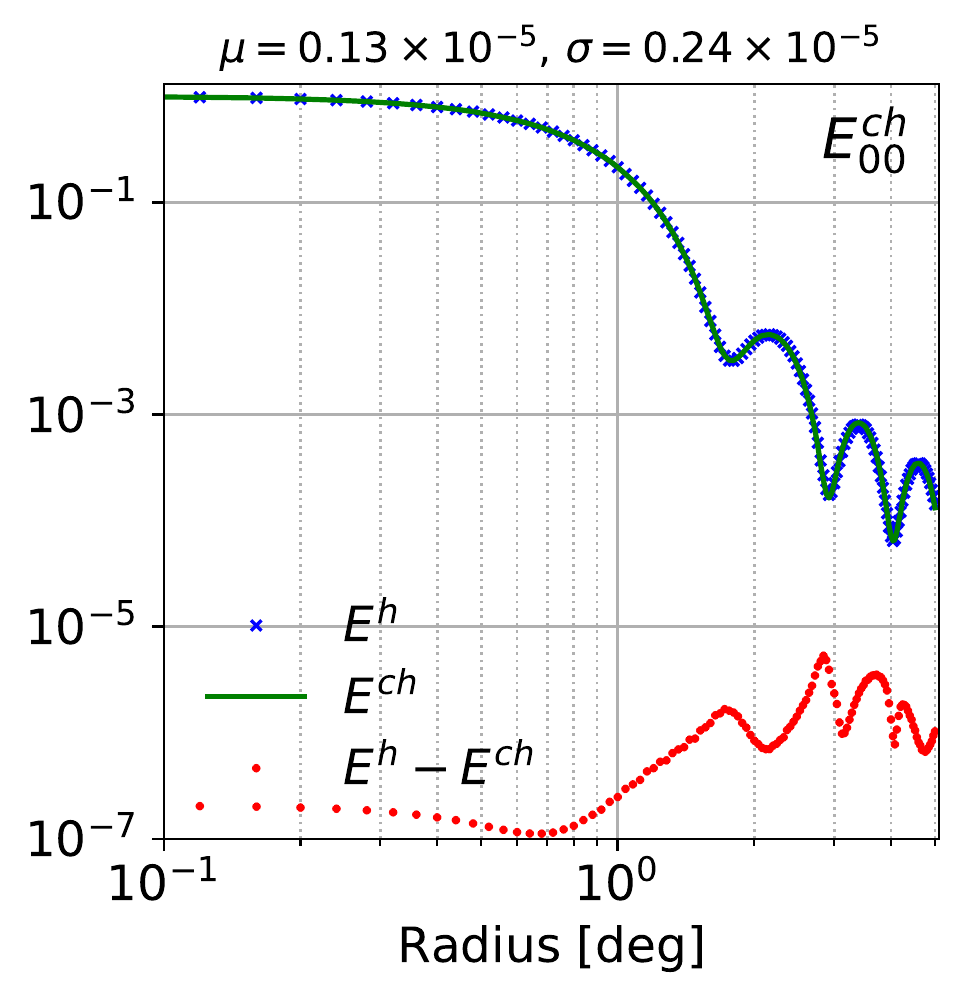}
	\includegraphics[width=0.24\linewidth]{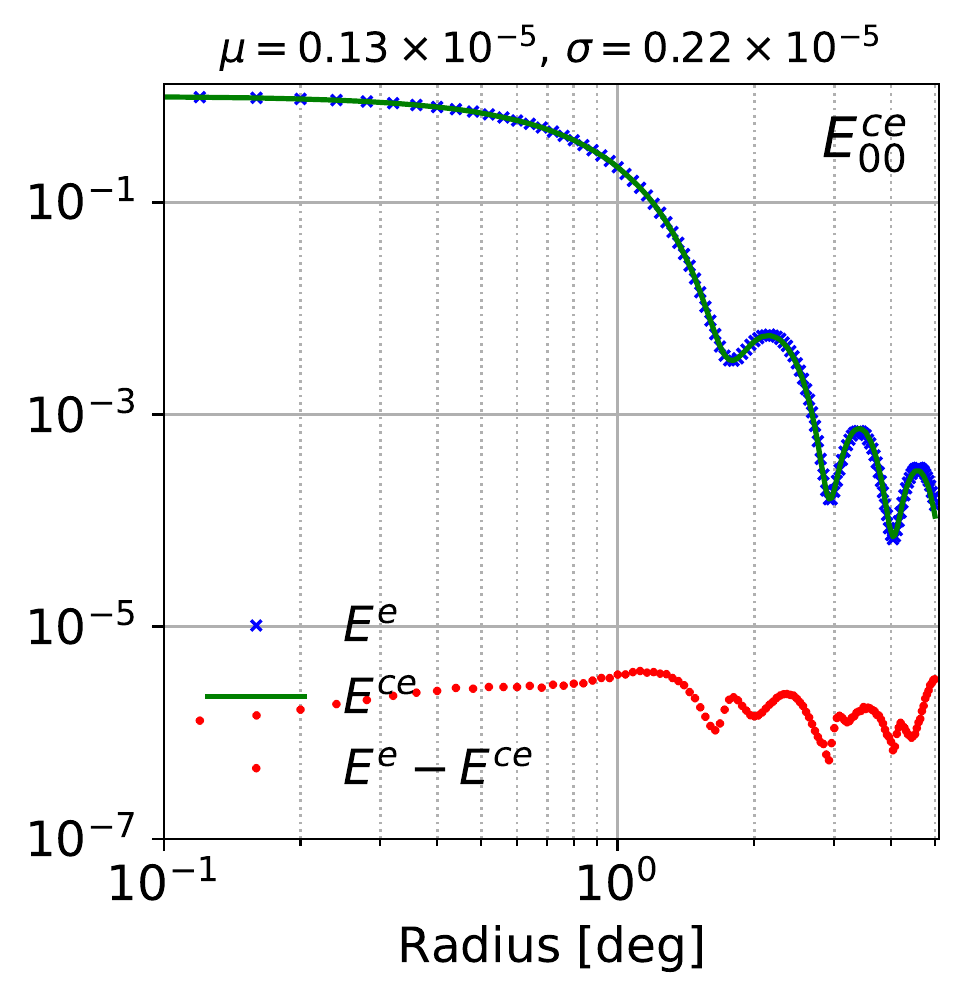}
	\includegraphics[width=0.24\linewidth]{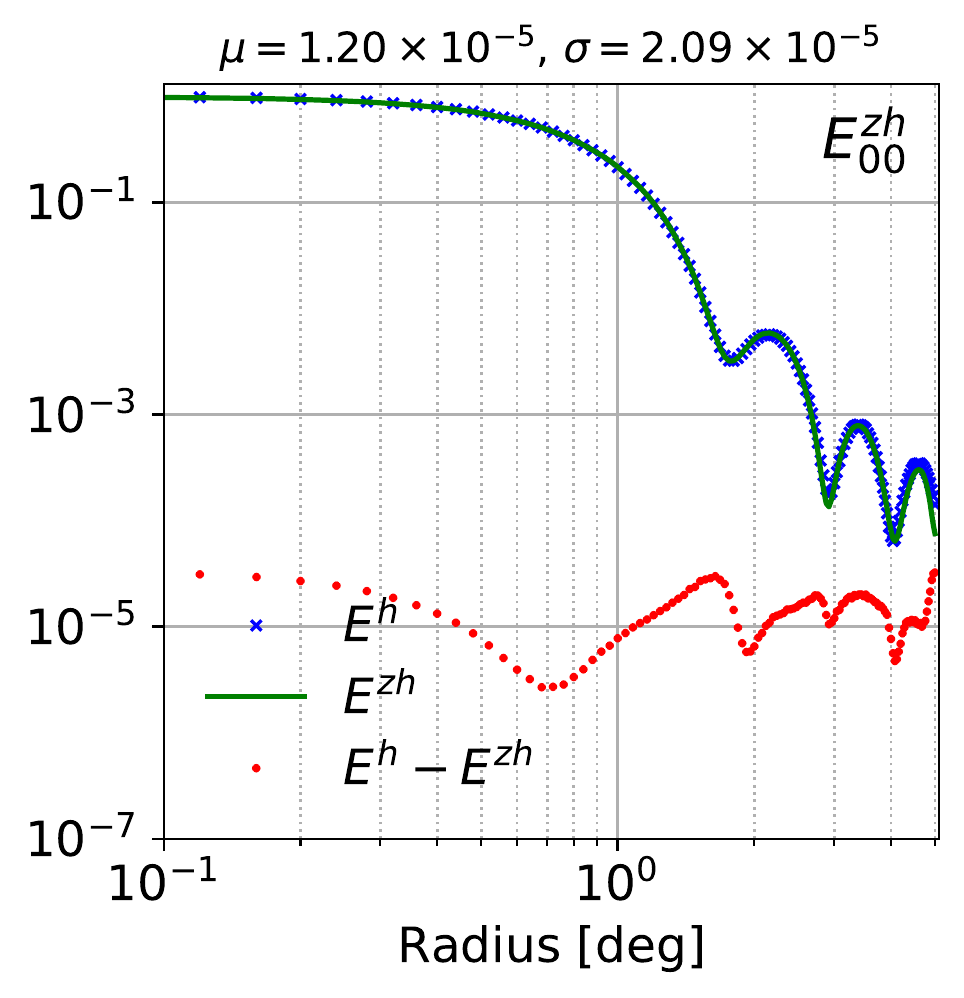}
	\includegraphics[width=0.24\linewidth]{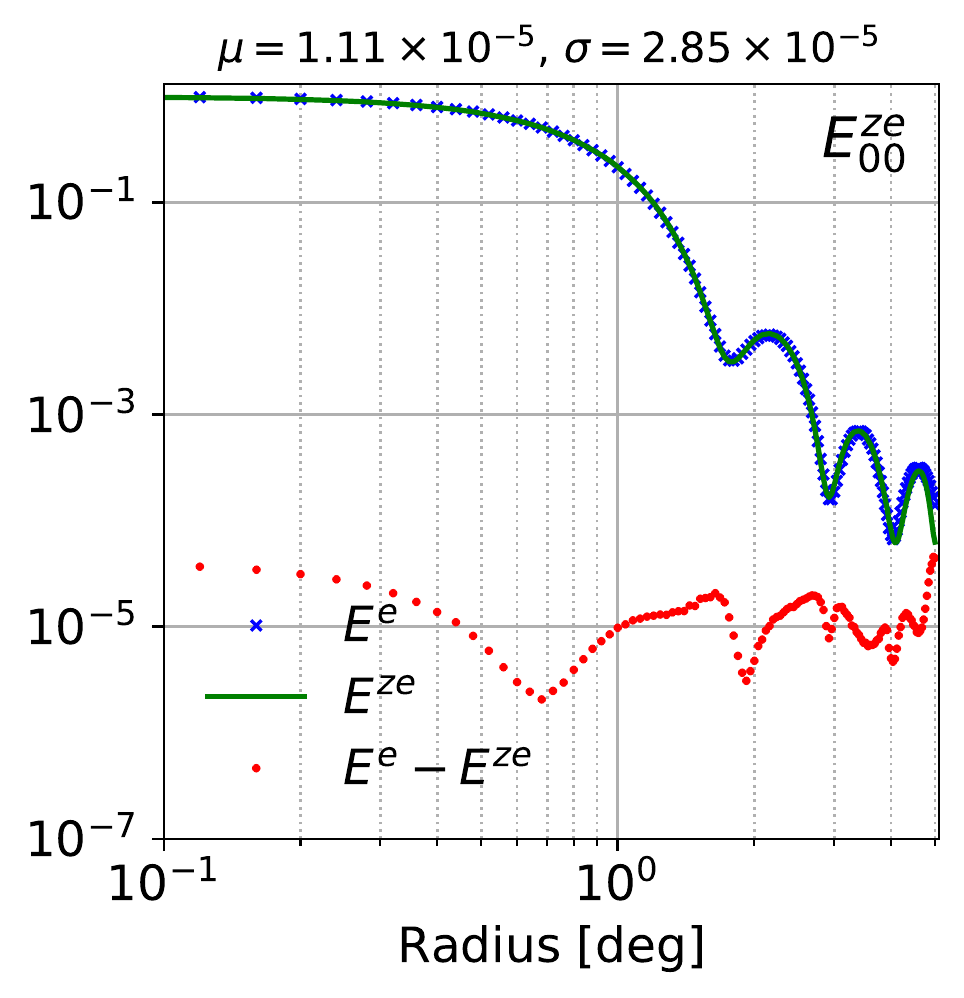}
	\caption{The principal component model created from the AH observation (left panels; $\mathbf{E}^{ch}$) and EM simulation (right panels; $\mathbf{E}^{ce}$) at 1070 MHz. The models and the residuals ($\mathbf{E}^{h}-\mathbf{E}^{ch}$ and $\mathbf{E}^{e}-\mathbf{E}^{ce}$) are shown in the top panels and the bottom panels show the radial profiles (averaged over concentric annuli) of the data, model and residual for the $E_{00}$ and $E_{01}$ elements. In the top panels, the power beams are shown in dB unit  and the residual images are multiplied by 100 to improve visualisation.}
	\label{f:Ec}
\end{figure*}

Furthermore, the proximity between the observation and simulation can be quantified as a fractional difference with the normalised root-mean-square error (NRMSE) of the magnitude of the EM model with respect to that of the given AH measurement.
We use NRMSE to show the overall (as the images are averaged over 10 degrees)  similarity of the two dataset as shown in Fig. \ref{f:nrmse-em}.
It is defined as
\begin{equation}
\text{NRMSE} = \frac{\sqrt{\overline{(|\mathbf{E}^h|-|\mathbf{E}^e|)^2}}}{\overline{|\mathbf{E}^h|}},
\end{equation}
as a function of frequency.
The NRMSE of the off-diagonal, i. e. cross-polarization, elements is, on average, around four times higher than that of the diagonal elements and the error increases with frequency.
Again, to what extent this discrepancy is due to actual error of the EM models cannot be known for certain because the low-level off-diagonal elements are less well-known in the AH observations and also become more noisy at higher frequencies.
The NRMSE of the diagonal elements increases smoothly with frequency, but that of the off-diagonal elements increases rapidly after around 1350 MHz.
The outliers in this figure are caused by Radio Frequency Interference (RFI) and not due to any intrinsic effect of the model or the measurement.
Note that the NRMSE is averaged over a diameter of 10 degrees and, hence, dominated by the errors near the nulls.

\subsection{Accuracy of the beams} \label{s:a2a}
The AH observation and EM simulation are completely independent methods.
The first derives from data while the other from solving differential equations.
Therefore, we can compare them to check their relative consistency at various levels of accuracy.
The EM simulation accounts for the mean features of the beam whereas AH observation helps representing the actual characteristics of the antennas.
The NRMSE of the diagonal elements of $\mathbf{E}^e$ is around 0.1 over 10 degrees which is very low considering the fact that there are as many as three nulls and sidelobes within this diameter; the error would be much lower within the main lobe. The good match is also evident in the radial profiles of the two beams shown in Fig. \ref{f:em}.
Therefore, we can safely say that the observation and simulation match well with each other for the diagonal elements.

The case is very different for the off-diagonal (or cross-polarization) elements.
The corresponding NRMSE is four times higher and can be more than 1.0 at higher frequencies.
As mentioned above, this discrepancy could be due to the fact that the polarized observation was not calibrated properly.
Due to the lack of available polarization-calibrated data and the lack of information regarding the accuracy of simulated cross-polarization beam, we are not considering the off-diagonal elements in our sparse beam model described in Section \ref{s:modelling}.

We use an AH beam averaged over three antennas, but the beam is known to vary from antenna to antenna.
The right panel of Fig. \ref{f:nrmse-em} shows the NRMSE of the beams of the three antennas with respect to the averaged beam $\mathbf{E}^h$ averaged over a diameter of 10 degrees for both the diagonal (below NRMSE = 0.1) and off-diagonal (above NRMSE = 0.1) elements.
For the diagonal elements, the antenna-to-antenna NRMSE variation range from 0.001 to 0.007 across the band which is almost an order of magnitude lower than the NRMSE of the simulated beam with respect to $\mathbf{E}^h$ (left panel).
M009 and M012 show similar variations across the band, but the variations of the M015 antenna is higher.
The variations in the off-diagonal terms are greater as expected and this gives us another reason for not trusting the polarization data at this stage.
Even though the NRMSE of the M015-beam is almost a factor of 2 higher than the other two antennas, we can consider the averaged beam to be a good approximation for all antennas in this case because the NRMSE for all three antennas is considerably low.
However, we caution the readers that we are talking about only three antennas and the scenario might change when all the antennas are considered.

In the following section, we present two different sparse models for the diagonal elements of the beam Jones matrix, one for the AH observation and the other for the EM simulation, both within a field of view of 10 degrees, and compare the two models at every stage.
A single model combining information from both observation and simulation might be beneficial in some cases, as discussed in the introduction, but we have kept the models separate for now.
Our main motivation behind modelling the two datasets in parallel was to compare the results of the noisy (observation) and noiseless (simulation) cases.

\section{Beam modelling} \label{s:modelling}

The squint-less, normalised, differential AH measurement $\mathbf{E}^h$ and the EM simulation $\mathbf{E}^e$ are effectively the starting point for the main body of our work.
We model the spatial beamshape using characteristic and analytic basis functions and the spectral behaviour using DCT.
This creates a sparse representation of the given beam datasets that can be used in direction dependent calibration and imaging effectively.

\subsection{Spatial modelling} \label{s:spat}

The beamshape is modelled using two types of orthonormal bases: `characteristic basis' (derived from PCA) and `analytic bases' (Zernike polynomials and spherical harmonics).
The method and results of the two approaches are described in the following two subsections.
Here, we will focus on just one frequency channel, centred at $\nu_c=1070$ MHz, because both $\mathbf{E}^h$ and $\mathbf{E}^e$ are sampled at this frequency, and the spectral behaviour of the spatial coefficients will be treated in the next section.

\subsubsection{Characteristic basis} \label{s:pca}
In this approach, the basis vector (eigenvector or eigenbeam) to describe a beam is created from the AH beam measurement itself, akin to the Characteristic Basis Function Pattern (CBFP) method \citep[e. g.][and references therein]{Maaskant2012,Mutonkole2016}.
Following the method described in section 4.3 of \textit{Paper I}, we have used Principal Component Analysis (PCA) to create such eigenbeams with the help of Singular Value Decomposition (SVD).

As opposed to the analytic bases, it is more appropriate to apply PCA on the spatio-spectral beam-cube, containing data for all the L-band frequency channels of $\mathbf{E}^h$ and $\mathbf{E}^e$.
We remove the RFI-affected channels from the AH measurement, stack $\mathbf{E}^h$ and $\mathbf{E}^e$ together and flatten the spatial dimensions to create a 4D array $\mathbf{E}^{he}$ of shape $N_\nu\times 2\times2\times N_{px}$, where $N_\nu=N^h_\nu+N^e_\nu$ is the total number of RFI-free frequency channels, and $N_{px}$ the total number of pixels in a beam image.
For each complex Jones element, the $N_\nu\times N_{px}$ matrix $\mathbf{E}^{he}(\nu)$ is decomposed into three complex matrices---$\mathbf{U}$, a $N_\nu\times N_\nu$ unitary matrix, $\mathbf{\Sigma}$, a $N_\nu\times N_\nu$ diagonal matrix whose diagonals contain all the singular values, and $\mathbf{V}^c$, a $N_\nu\times N_{px}$ unitary matrix---using SVD.\footnote{The python module {\tt scipy.linalg.svd} is used.}
Here, $\mathbf{V}^c$ is the collection of our basis vectors or principal components (PC), i. e. eigenbeams, and $\mathbf{C}^c=\mathbf{U} \mathbf{\Sigma}$ contains the corresponding coefficients \citep[for the relationship between SVD and PCA, see][section 2a]{Jolliffe2015}.
The models for all frequencies of L-band are reconstructed from the eigenbeams and the coefficients as $\mathbf{E}^c(\nu)=\mathbf{C}^c_0 \mathbf{V}^c_0$, where $\mathbf{C}^c_0$ contains the strongest coefficients and $\mathbf{V}^c_0$ the corresponding PCs.

\begin{figure}
\includegraphics[width=\columnwidth]{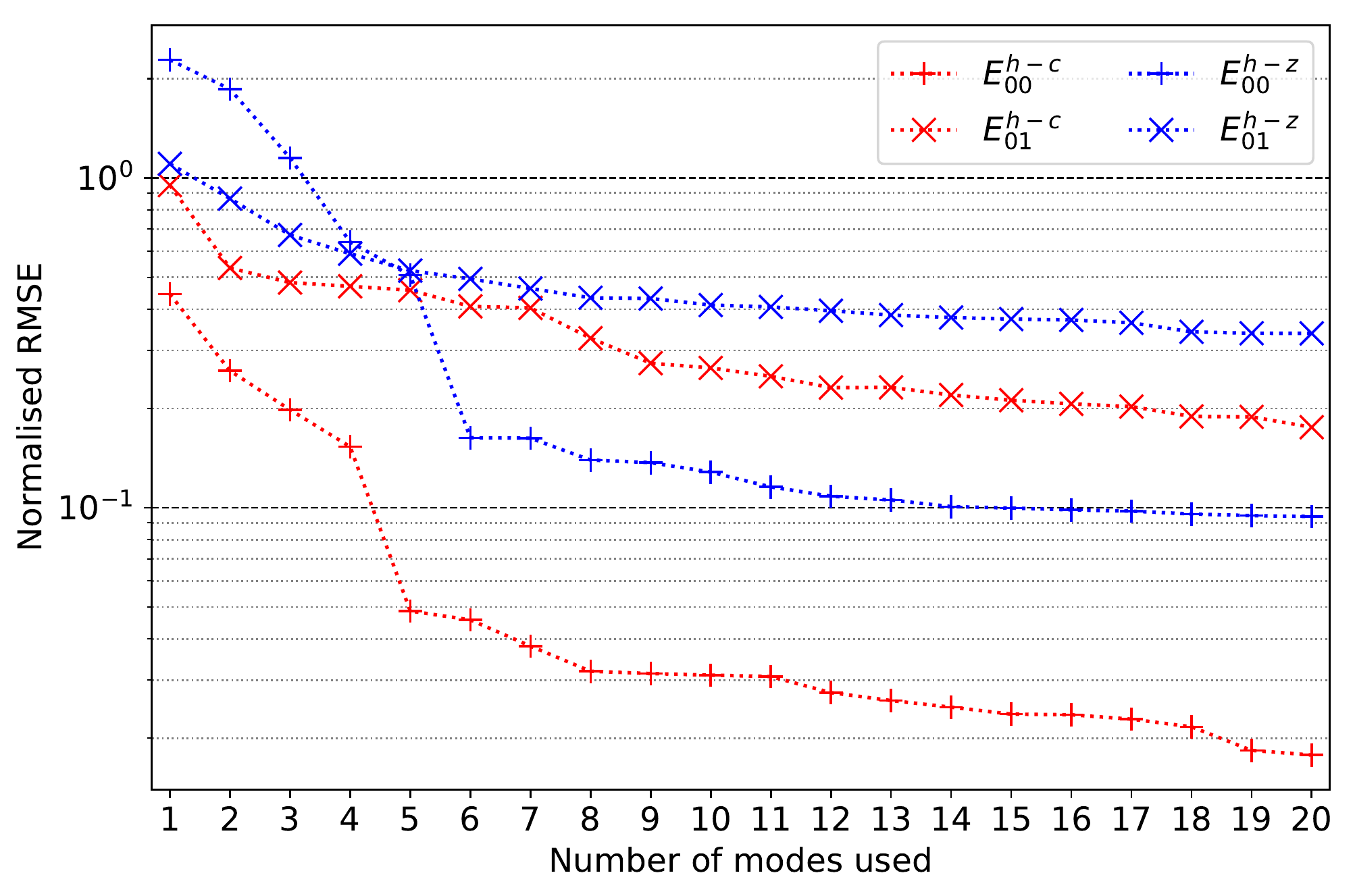}
\caption{Normalised root-mean-square error (NRMSE) of the principal component model (red) and the Zernike model (blue) as a function of the number of modes used at a frequency of 1070 MHz averaged over a diameter of 10 degrees. The plus and cross signs represent the errors for the $E_{00}$ and $E_{01}$ elements, respectively.}
\label{f:nrmse}
\end{figure}

The results over the full L-band will be presented in Section \ref{s:nu}; here we focus on a single channel.
The top row of Fig. \ref{f:Ec} shows the reconstructed diagonal Jones elements of $\mathbf{E}^{ch}(\nu_c)$ (first from the left; AH) and $\mathbf{E}^{ce}(\nu_c)$ (second from the left; EM) where $\nu_c=1070$ MHz.\footnote{In this paper, $\mathbf{E}^x$ refers to a model created using the basis function $x=\{c,z\}$ for both AH observation and EM simulation whereas $\mathbf{E}^{xh}$ and $\mathbf{E}^{xe}$ refer to the AH and EM models created using the basis, individually.}
The corresponding residuals after subtracting the models from $\mathbf{E}^h$ are shown below the models.
Note that all the residual images are calculated from the actual amplitudes of the Jones elements, not the squared amplitudes, i. e. power.
These models are created using $N^c=15$ strongest PCs.
The NRMSE ($[\overline{(|\mathbf{E}^h|-|\mathbf{E}^{ch}|)^2}]^{1/2}/\overline{|\mathbf{E}^h|}$) of the model created from AH is shown in Fig. \ref{f:nrmse} as a function of number of modes used.
We see that PCA can represent the beam measurements accurately with less than 15 components (red plus markers), but we have used 15 modes for a fair comparison with the models created using analytic bases.

The bottom panels of Fig. \ref{f:Ec} show the radial profiles of the $E_{00}$ and $E_{11}$ Jones elements of the data (blue dots), the characteristic models (green line) and the residuals (red dots).
Radial profiles are created by averaging the beam images in concentric circular annuli.
In both the images of the top panels and the radial profiles of the bottom panels, we show the squared amplitude for the sake of consistency.
The mean and standard deviation of the residuals are shown above the radial profile plots.
If we take the square-root of these values, we see that with 15 PCs, we can reach an rms residual level of $\sim 10^{-3}$ for the diagonal Jones elements for both the AH and EM beams.

Along with a faithful representation of the beam, one other advantage of using an orthogonal basis is the sparsity of the representation---the beam can be described using less information.
The AH or EM datasets contained $4N_{px}\times N_\nu$ parameters for describing the full-Jones beam and although the characteristic basis model $\mathbf{E}^{c}$ now needs $4N^c(N_{px}+N_\nu)$ parameters to represent the full-bandwidth full-polarization beam, the frequency behaviour of the PC coefficients can be modelled using $N^c_\nu<N_\nu$ parameters, as described in Sec. \ref{s:nu}, further reducing the information-load.
However, $N_{px}$ will still be a large number, especially if a bigger field of view or better resolution is desired.
This is where the `analytic basis' approach comes in.
It can further sparsify the model by representing the spatial dimensions via 2D analytic functions calculated over unit disks.

\subsubsection{Analytic basis}
We test two types of analytic bases to represent $\mathbf{E}^h$ and $\mathbf{E}^e$---Zernike polynomials \citep[ZP; for an introduction, see][and Appendix \ref{ap:ZP}]{Lakshmi2011} and spherical harmonics (SH).
Unlike PCA, the analytic basis models are created for each AH and EM frequency channel independently.
The basic procedure is the same for both ZP and SH.
If the analytic bases are denoted by $\mathbf{V}=\{\mathbf{Z},\mathbf{S}\}$ for ZP and SH, respectively, and the corresponding coefficients by $\mathbf{C}$, then the decomposition of $\mathbf{E}^h$ gives us the coefficients
\begin{equation}
\mathbf{C} = (\mathbf{V}^T\mathbf{V})^+ \ \mathbf{V}^T \mathbf{E}^h
\end{equation}
where $^T$ stands for transpose and $^+$ for the Moore-Penrose pseudoinverse.\footnote{Calculated using the python module {\tt numpy.linalg.pinv}.} These coefficients can be used to reconstruct a beam model as
\begin{equation}
\mathbf{E}^m = \sum\limits_{i=0}^{N^m} \mathbf{C}_i \mathbf{V}_i
\end{equation}
where $i=0$ denotes the strongest among the selected coefficients, and $i=N^m$ the weakest, and $m=\{z,s\}$ for ZP and SH.
The results of the decomposition and reconstruction of the AH and EM datasets using these bases are presented below.

\begin{figure}
	\centering
	\includegraphics[width=0.47\columnwidth]{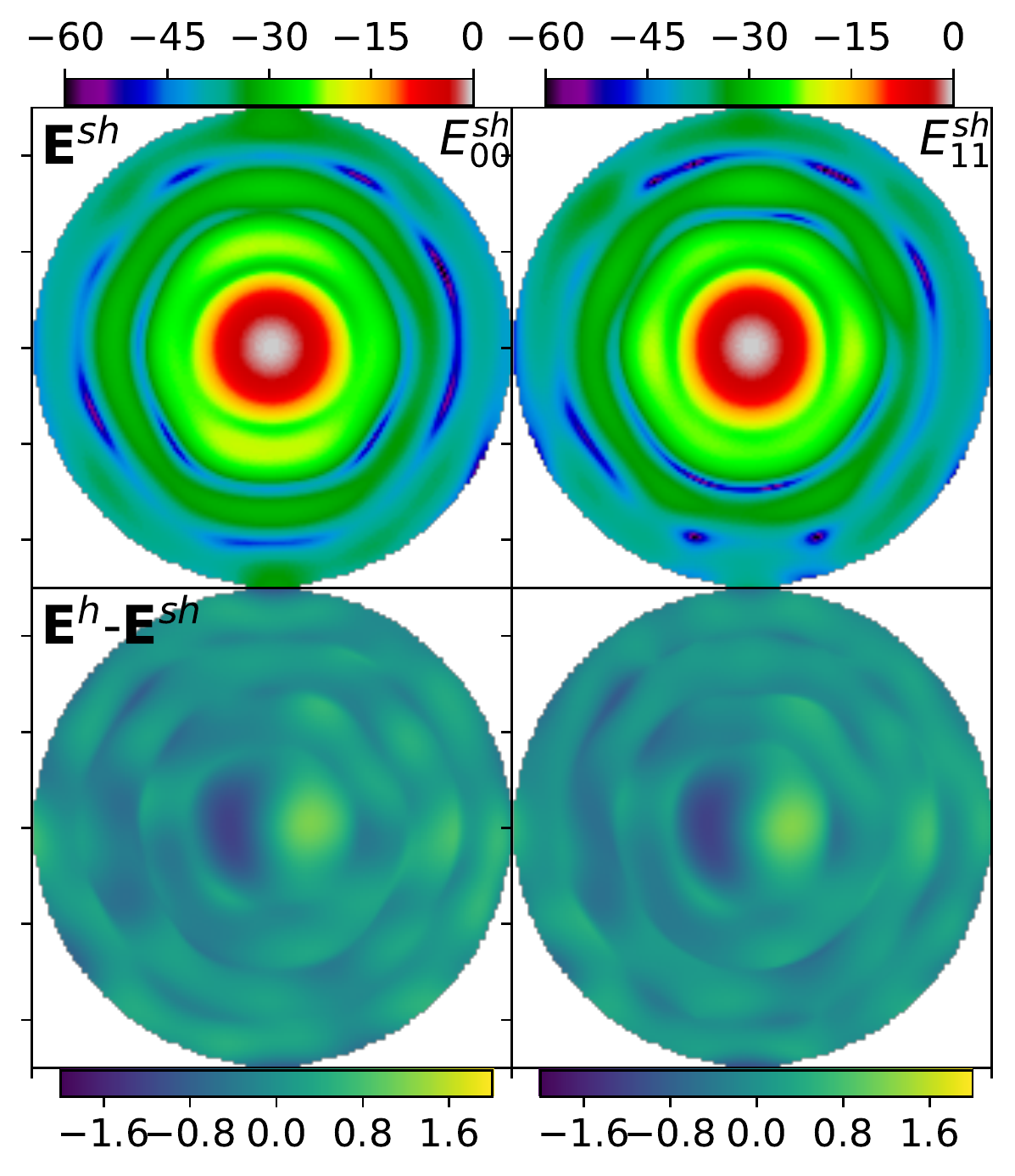}
	\includegraphics[width=0.51\columnwidth]{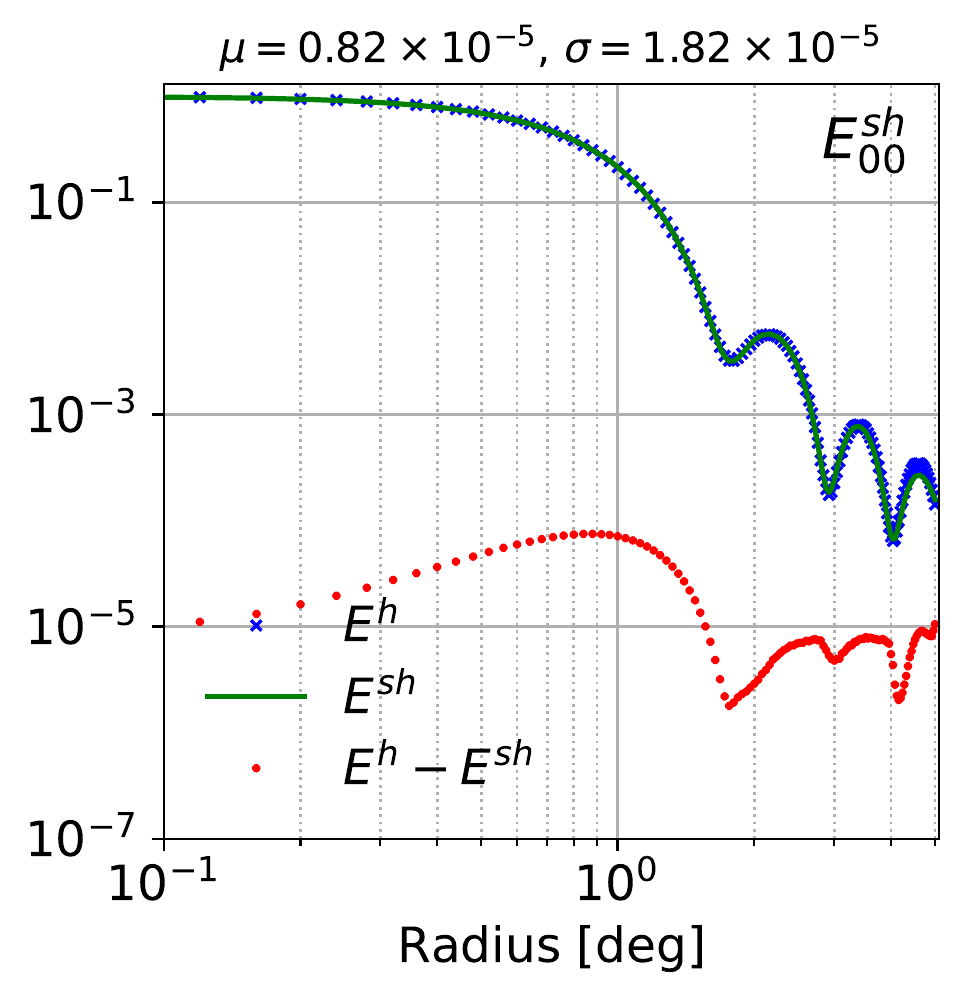}
	\caption{The beam model created by decomposing AH observations using spherical harmonics. We have used 40 and 15 modes for the diagonal and off-diagonal elements, respectively. The bottom panels show the corresponding residuals, multiplied by 100 for better visualisation.}
	\label{f:Es}
\end{figure}

\subparagraph{Zernike polynomial model:}
ZP models $\mathbf{E}^{z}$ created from the AH ($\mathbf{E}^{zh}$) and EM ($\mathbf{E}^{ze}$) datasets are shown in the top panels of Fig. \ref{f:Ec}, the former third from the left and the latter last from the left.
Like the PCA model, the 15 strongest modes were used to reconstruct these.
We keep using 15 coefficients because, as shown in Fig. \ref{f:nrmse} and described in more detail below, modelling error cannot be reduced substantially by including more coefficients in the case of a single frequency channel.
By comparing the residual images of the two leftmost panels with those of the two rightmost panels of Fig. \ref{f:Ec}, we immediately see that PCs can represent the data more faithfully with the same number of modes.
A comparison of the corresponding radial profiles reveals that the residual levels for $\mathbf{E}^{zh}$ and $\mathbf{E}^{ze}$ are, on average, an order of magnitude higher than those for $\mathbf{E}^{ch}$ and $\mathbf{E}^{ce}$.

NRMSE of $\mathbf{E}^{h-z}$ (difference between AH measurement and ZP representation) is shown in Fig. \ref{f:nrmse} (blue plus markers for $E_{00}$ and blue crosses for $E_{01}$ for comparison) as a function of number of modes used; $\mathbf{E}^{e-z}$, not shown here, follows a similar trend.
After using the 15 strongest Zernike modes, the NRMSE of $E_{00}$ is $\sim 0.1$ and that of $E_{01}$ around 0.3;
compare them to the corresponding NRMSE for $\mathbf{E}^{h-c}$ (red plus and cross)---around 0.02 and 0.2.
Comparing the blue and red markers, we see that the NRMSE for $\mathbf{E}_{00}^{h-z}$ and $\mathbf{E}_{00}^{h-c}$ decreases in a similar fashion as more modes are included, but they always maintain a difference of almost an order of magnitude.
The difference is lower for the off-diagonal elements as both of them are dominated by noise; we do not include these cross-polarization terms in our final model.

ZP are real functions, therefore, we model both the real and imaginary parts using the real functions and store the resulting complex coefficients.
The beam measurement and simulation were decomposed using the first 300 Zernike modes and the 15 strongest ones among them were selected for this particular reconstruction at 1070 MHz.
As discussed before, the main reason for using analytic bases is to reduce the amount of information needed to represent the spatial structure.
It is, therefore, imperative to look at the shapes of the Zernike modes needed for that representation.
However, we want to find the strongest Zernike coefficients needed to model both the beam measurements and simulations for all frequencies and, hence, the shapes of the individual modes is discussed in Section \ref{s:nu}.

\begin{figure*}
	\centering
	\includegraphics[width=\linewidth]{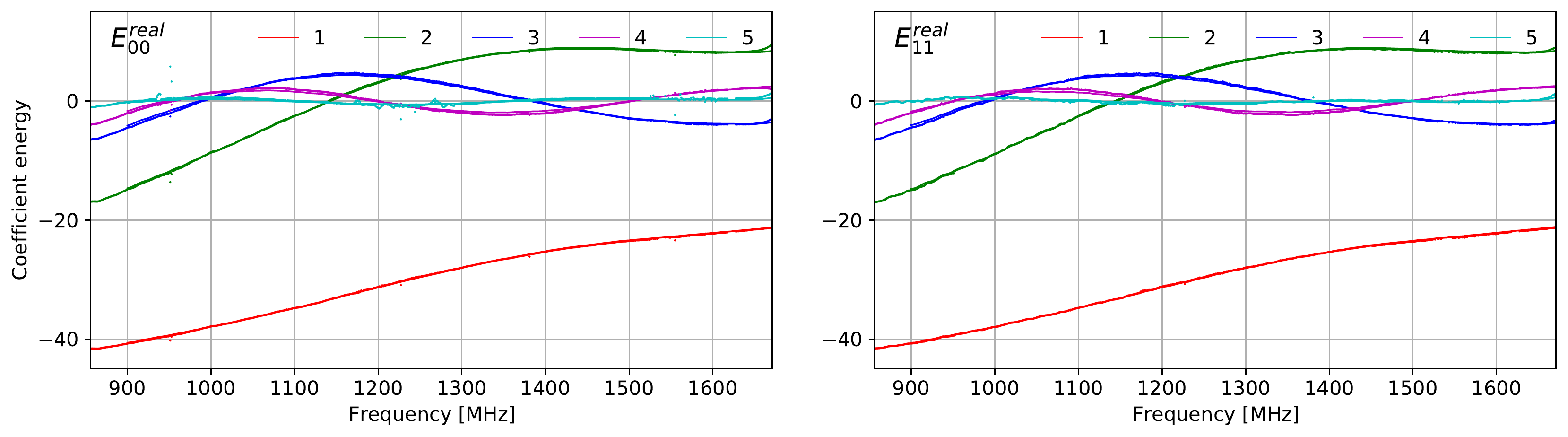}
	\includegraphics[width=\linewidth]{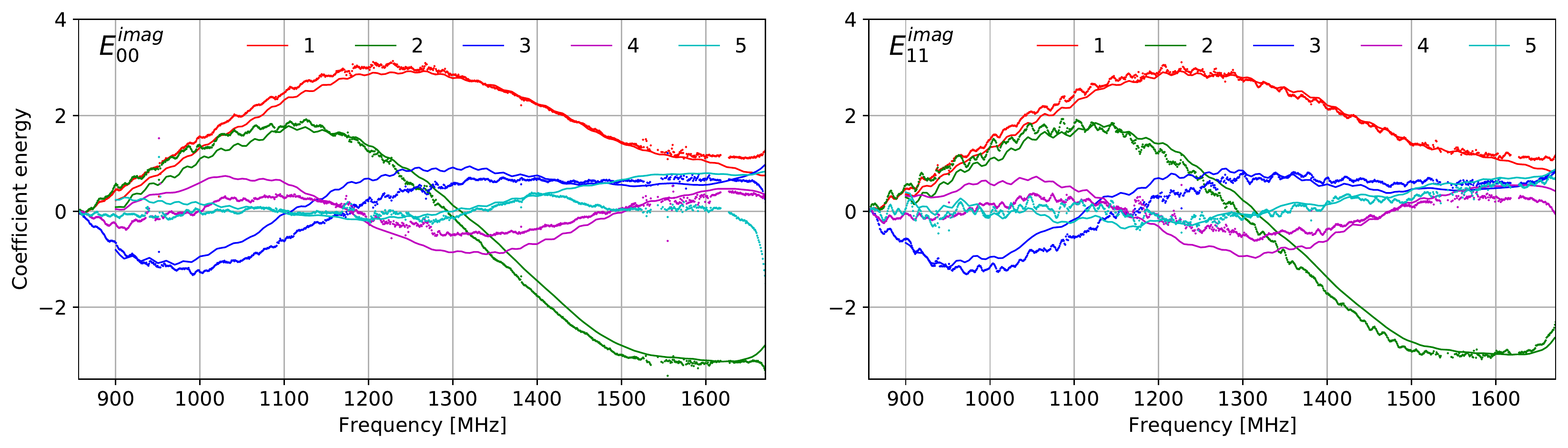}
	\caption{The strongest principle components, as a function of frequency, for the AH (dots) and EM (solid lines) models. The top two panels show the real parts of the diagonal Jones elements and the bottom two panels the imaginary parts.}
	\label{f:s-pca}
\end{figure*}
\begin{figure*}
	\centering
	\includegraphics[width=\linewidth]{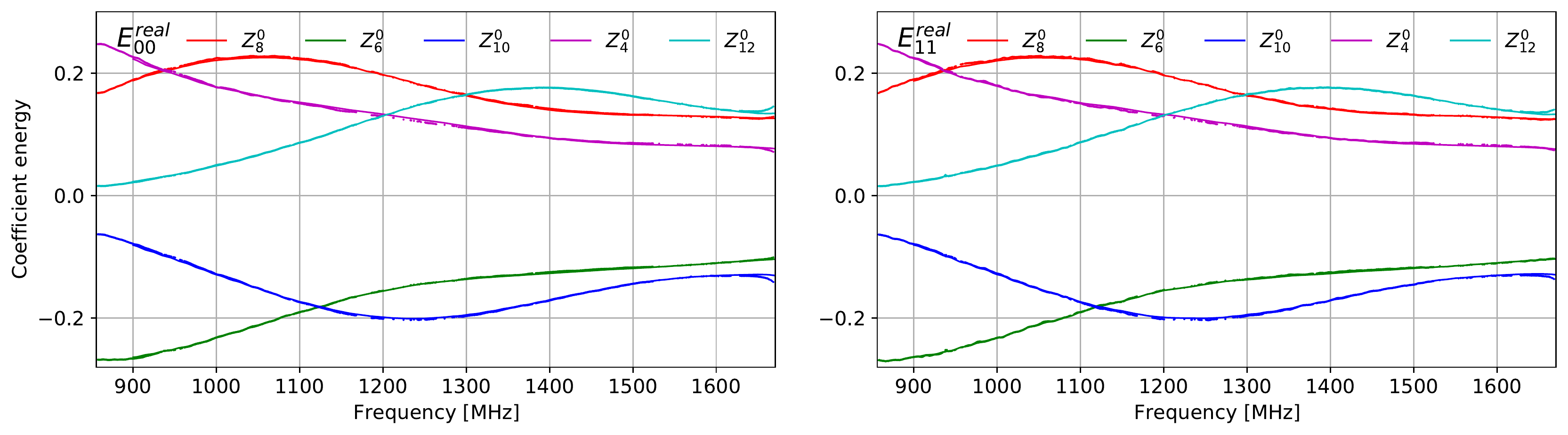}
	\includegraphics[width=\linewidth]{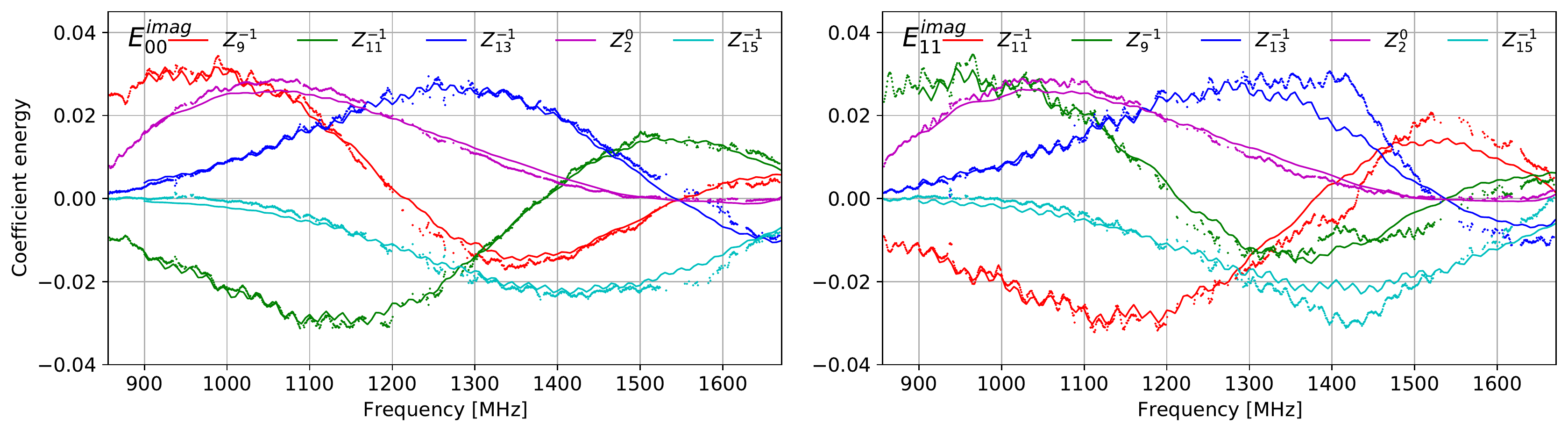}
	\caption{The five strongest Zernike coefficients, as a function of frequency, for the AH (dots) and EM (solid lines) models. The top two panels show the real parts of the diagonal Jones elements and the bottom two panels the imaginary parts.}
	\label{f:s-zern}
\end{figure*}

\subparagraph{Spherical harmonic model:}
Unlike the ZP, the SH are complex valued although their imaginary parts do not have any zero-frequency component and for any order (or degree) $n$, the frequency $m=0$ contains the average of all the harmonics.
We decompose the observed complex beam $\mathbf{E}^h$ using the first 256 complex SH (with a maximum $n$ of 15) and select the strongest coefficients.
Unlike PCs and ZPs, SH's cannot reconstruct the diagonal elements accurately with 15 modes and at least 40 modes are needed for a reasonable representation.
Therefore, the the SH model $\mathbf{E}^{sh}$, shown in Fig. \ref{f:Es} (left panel), has been created using 40 coefficients.
Even with 40 modes, SH does not perform as well as PC and ZP---the residuals show a dipolar structure in the main lobe of the beam.
This precludes using SH for modelling 10-degree beams across a wide bandwidth and, hence, in the next section, we will focus only on ZP and PC.

\subsection{Spectral modelling} \label{s:nu}
The two full-bandwidth sparse representations $\mathbf{E}^c$, based on PC, and $\mathbf{E}^z$, based on ZP, can be compressed even further if we model the spectral behaviour of the associated strongest coefficients in each case.
PCs are calculated from the combined dataset of the full-bandwidth AH and EM beams, but ZP decomposition is performed at each frequency channel separately.
To prepare a dataset for PCA, we removed the RFI-affected channels from AH measurements, as mentioned in Section \ref{s:pca}, and created a new dataset containing only the non-contaminated AH channels and all the EM channels.

We used three different figures of merit to detect RFI-affected channels in the AH measurement: the rms of the beam images, the position of the peak in the diagonal elements and the difference between the energies of the ZP coefficients of adjacent channels.
In the RFI-affected channels, rms is comparatively high and the beam is not exactly centred at the central pixel (128,128).
The dataset created after removing the unusable channels based on these two criteria was used for PCA.
However, these criteria could detect only the worst channels.
To identify the remaining lower-level RFI, we used ZP coefficients $\mathbf{C}^{zh}(\nu)$.
ZP decomposition was performed on all channels including the RFI-affected ones because, unlike PC, ZP coefficients of one channel do not depend on those of another.
Then, we identified the channels $\nu_i$ for which $C^{zh}_{00}(\nu_i)-C^{zh}_{00}(\nu_{i+1}) > 10^{-3}$ as RFI-affected.
Finally, we created a comprehensive list of RFI-affected channels and masked those channels in both $\mathbf{C}^{zh}(\nu)$ and the PC coefficients $\mathbf{C}^{ch}(\nu)$.

\begin{figure*}
	\centering
	\includegraphics[width=\linewidth]{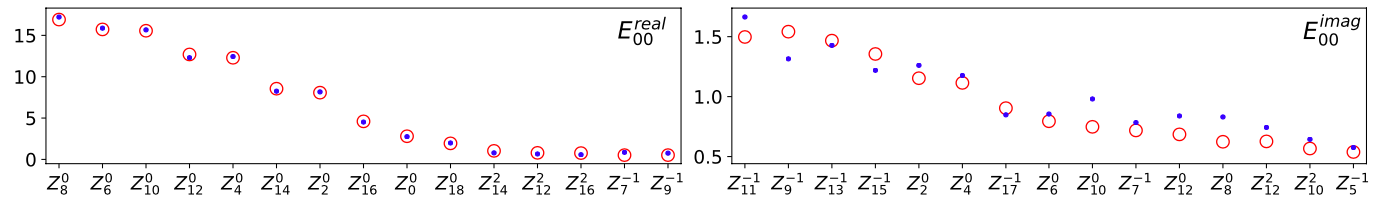}
	\caption{Frequency-averaged energies of the 15 strongest Zernike coefficients for the AH (red) and EM (blue) models. The energies are multiplied by 100 for better visualisation. The coefficients are selected from the spectral average of the AH and EM models together, but the corresponding energies are shown separately.}
	\label{f:zern}
\end{figure*}

\subsubsection{Principal components}
The most dominant PCs are shown in Fig. \ref{f:s-pca} as a function of frequency for the diagonal Jones elements of $\mathbf{E}^{ch}$ (dots) and $\mathbf{E}^{ce}$ (solid lines).
AH measurements are available from 856 to 1712 MHz with a resolution of 0.83 MHz, but we show results up to the effective L-band edge, 1670 MHz.
EM simulations are available from 900 to 1670 MHz with a resolution of 5 MHz.
The top two panels in the figure show the real parts of $\mathbf{C}^{c}(\nu)$ and the bottom ones the imaginary parts.
The first 5 real PCs of the diagonal elements of $\mathbf{E}^h$ and $\mathbf{E}^e$ match very well and their difference is only visible in the imaginary parts which are at a much lower level.

The amplitude of the strongest PC (red lines in Fig. \ref{f:s-pca}) models the beamwidth as a function of frequency to a large extent.
This amplitude decreases smoothly with frequency because beamwidth is proportional to $\lambda/D$, but it also exhibits a low-level frequency-dependent ripple, more clearly visible in the imarinary part (red lines in $E_{00}^{imag}$ and $E_{11}^{imag}$).
The ripple of MeerKAT beam is described in more detail in Section \ref{s:ripple}.

\subsubsection{Zernike coefficients}
In order to select the most dominant Zernike coefficients $\mathbf{C}^{z}(\nu)$, we calculate the average of the absolute value of the real and imaginary parts separately over all frequencies for both AH measurement and EM simulation.
The 15 strongest Zernike modes selected from these datasets is shown in Fig. \ref{f:zern}.
The red circles and blue dots show the average energies of the coefficients for $\mathbf{E}^h$ and $\mathbf{E}^e$, respectively.
We see that the real part of the diagonal elements is mainly represented by the Zernike modes with an angular frequency $m=0$, i. e. the spherical modes.
The spherical modes are symmetric, but the beam has asymmetries which are represented by the astigmatism ($m=2$) and coma ($m=-1$) modes.
The imaginary part is also represented by the spherical, astigmatism and coma modes, but astigmatism is much more dominant than the spherical modes.
The energies of $E^h_{00}$ and $E^e_{00}$ match very well in the real part as opposed to the imaginary part.

Spectral behaviour of the most dominant Zernike coefficients is presented in Fig. \ref{f:s-zern}.
Like Fig. \ref{f:s-pca}, $\mathbf{C}^{zh}$ are plotted using dots and $\mathbf{C}^{ze}$ using solid lines.
Similar to the PCs, the strongest Zernike coefficients for modelling the real part of the diagonal elements match very well between $\mathbf{E}^h$ and $\mathbf{E}^e$ and the ripple of the beamwidth is also clearly visible in the imaginary parts.
Note that higher order spherical modes are needed for modelling higher frequency beams because number of sidelobes within the $10^\circ$ diameter increases with frequncy.\\

The spectral shape of the spatial coefficients is modelled using DCT as described below.
We discuss spectral compression only for the ZP case because this is the basis we have used in our final models and ZPs can represent the beam using less information.
The spectral behaviour of the PCs can be modelled using the same method if needed.

\begin{figure*}
	\centering
	\includegraphics[width=\linewidth]{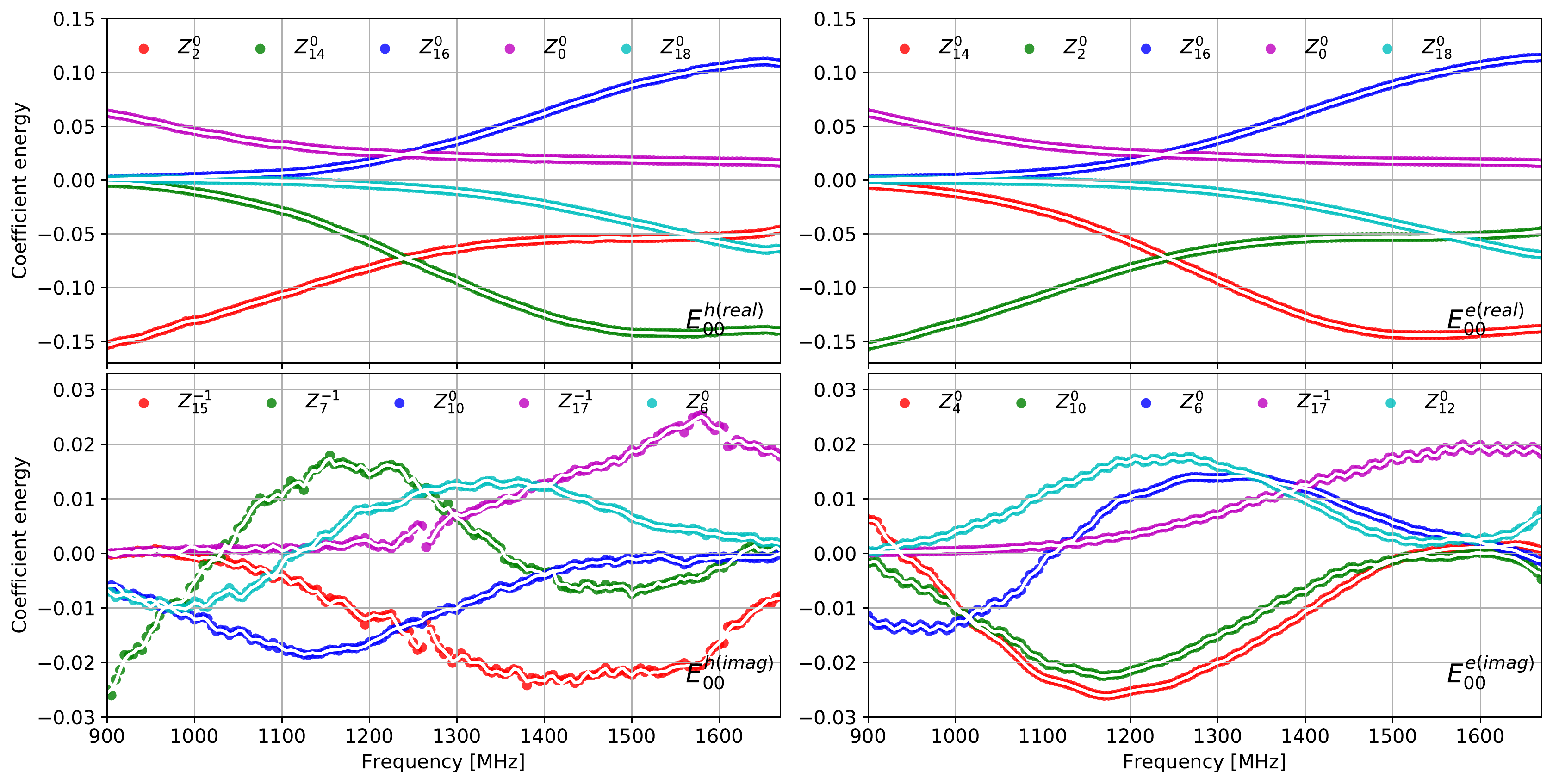}
	\caption{Spectral modelling of the sixth to tenth strongest Zernike coefficients using DCT for AH (left panels) and EM models (right panels). Both the real (top panels) and imaginary (bottom panels) parts are shown. The thick coloured lines show the original interpolated energies of the coefficients and the thin white lines running through them represent the corresponding DCT reconstructions.}
	\label{f:dct}
\end{figure*}

\begin{figure*}
	\centering
	\includegraphics[width=\linewidth]{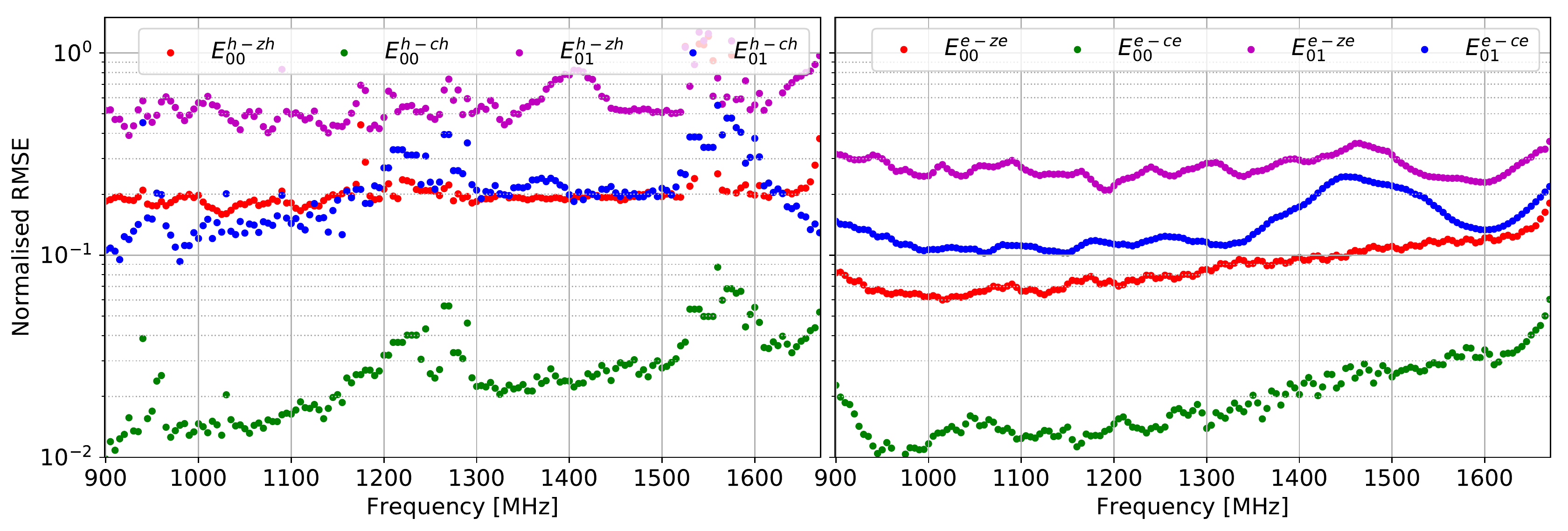}
	\caption{Normalised root-mean-square error of the ZP and PC-based AH and EM models as a function of frequency. Left panel shows the error of $\mathbf{E}^{zh}$ and $\mathbf{E}^{ch}$ with respect to the given AH measurement and the right panel the error of $\mathbf{E}^{ze}$ $\mathbf{E}^{ce}$ with respect to the given EM simulation.}
	\label{f:mse}
\end{figure*}

\begin{figure}
	\centering
	\includegraphics[width=\columnwidth]{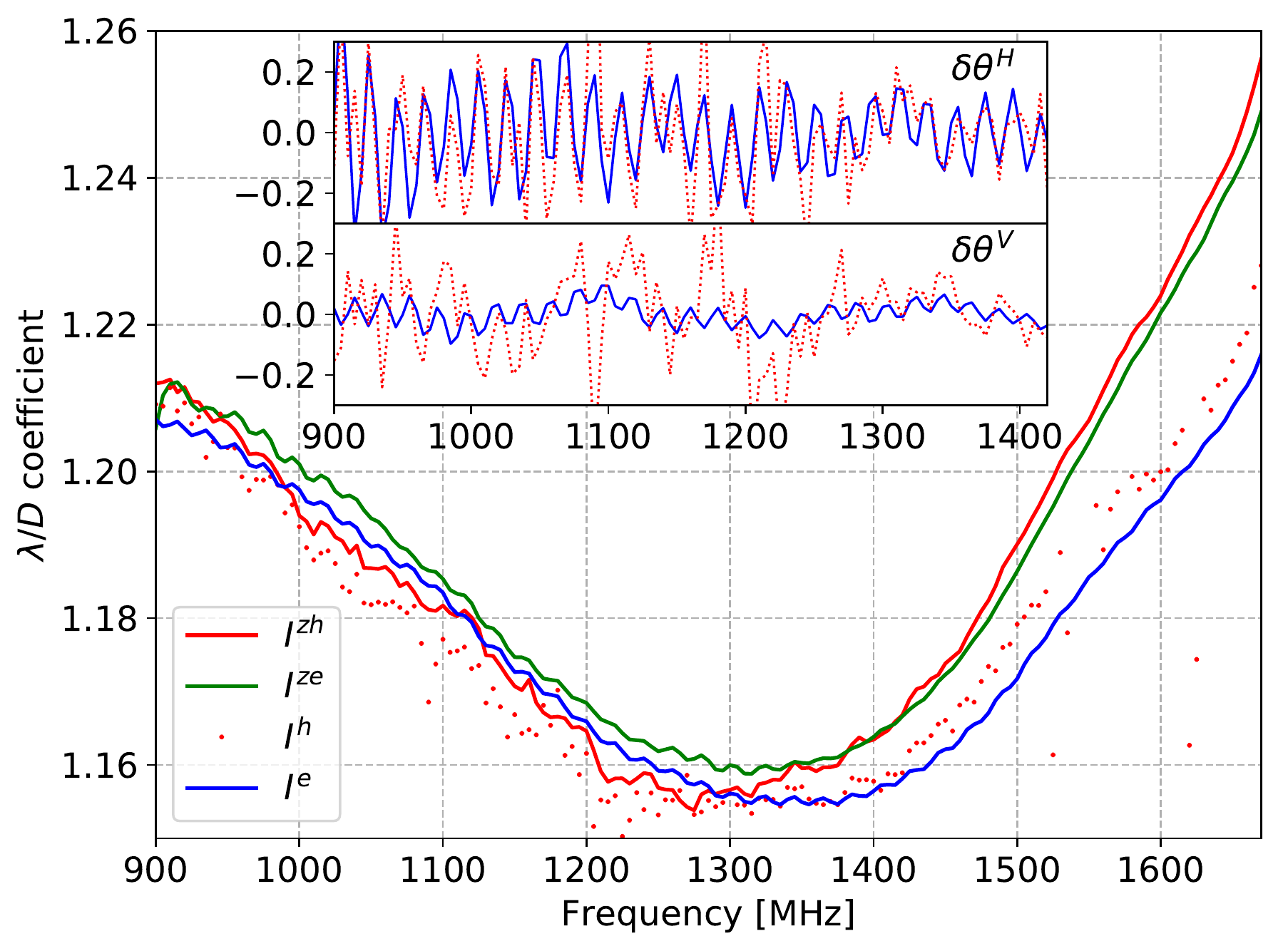}
	\caption{Coefficients of the theoretical beamwidth ($\lambda/D$) as a function of frequency for the given AH (red dots) and EM (blue) datasets, and the Zernike-based models created from the AH (red solid line) and EM (green) datasets. The inset plot shows the amplitude of the beamwidth ripple in arcmin units in the horizontal ($\delta\theta^H$) and vertical ($\delta\theta^V$) directions for the AH (red) and EM (blue) datasets.}
	\label{f:ripple}
\end{figure}

\subsubsection{Discrete cosine transform} \label{s:dct}
Only 15 ZP coefficients were needed to model the beam at 1070 MHz as shown in Fig. \ref{f:Ec} and \ref{f:nrmse}, but at least 20 coefficients are necessary for the full L-band.
Therefore, we model the spectral behaviour of the 20 most dominant spectral coefficients $\mathbf{C}^{z}$, i. e. $\mathbf{C}^{zh}$ for AH and $\mathbf{C}^{ze}$ for EM.
Before compressing the coefficients, we interpolate\footnote{The python module {\tt numpy.interp} is used to perform a piecewise linear interpolation.} their energies, for both AH and EM models, to $N^i_\nu=7701$ channels from 900 to 1670 MHz with a resolution of 0.1 MHz.
In case of $\mathbf{C}^{zh}$, the coefficients for the RFI-affected channels were reconstructed by interpolation from the RFI-free ones and then the coefficients of $N^h_\nu=1024$ channels were used to fill $N^i_\nu$ channels via piecewise linear interpolation.
In case of $\mathbf{C}^{ze}$, coefficients from the original $N^e_\nu=155$ channels are used to fill the $N^i_\nu$ channels.

Interpolated coefficients $\mathbf{C}^{z}$ are then compressed via DCT.
First, we decompose the coefficients using DCT of Type II as
\begin{equation}
\mathbf{C}_k = \frac{2}{\sqrt{wN^i_\nu}} \sum\limits_{n=0}^{N^i_\nu-1} \mathbf{C}_n(\nu) \cos\left[\frac{\pi k(2n+1)}{2N^i_\nu}\right]
\end{equation}
for $0\le k < N^i_\nu$ where $w=4$ when $k=0$ and $w=2$ otherwise.
We have seen that the number of resulting DCT coefficients that are more than 10 times higher than the noise level in the decomposition is usually around 30 for $\mathbf{C}^{ze}$ and 40 for $\mathbf{C}^{zh}$.
Therefore, we decided to keep 40 DCT coefficients for each Jones elements of the AH model and 30 for the EM model.
$\mathbf{C}^{zh}$ needs more coefficients because it has noise.

If the resulting most dominant DCT coefficients are denoted by $\mathbf{C}'$, we can reconstruct a de-noised smooth spectral model of the coefficients via Inverse DCT (same as DCT of Type III) as
\begin{equation}
\widehat{\mathbf{C}}_n(\nu) = \frac{\mathbf{C}'_0}{N^i_\nu}+\sqrt{\frac{2}{N}} \sum\limits_{k=0}^{N^i_\nu-1} \mathbf{C}'_k \cos\left[\frac{\pi n}{N^i_\nu}\left(n+\frac{1}{2}\right)\right]
\end{equation}
for $0\le n < N_\nu$.
Both $\mathbf{C}^z(\nu)$ (thick coloured lines) and $\widehat{\mathbf{C}}^z(\nu)$ (thin white lines) are shown in Fig. \ref{f:dct} for both the AH (left panels) and EM (right panels) models.
The sixth to tenth most dominant coefficients for modelling $E^h_{00}$ and $E^e_{00}$ are shown because the first five coefficients, already shown in Fig. \ref{f:s-zern}, have even less error and are not representative of the rest of the coefficients.

From Fig. \ref{f:dct}, we see that the smooth reconstruction of the spectral behaviour follows the original energies of the coefficients closely.
DCT can also reconstruct the small-scale ripple on the coefficients to some extent, but further work is needed to model the ripple more accurately based on physical considerations.

We need only $2\times 2\times 2\times N^d\times N^z$ coefficients to represent the full L-band complex beam model of MeerKAT where $N^d$ is the number of DCT coefficients for modelling the spectral behaviour of each the $N^z$ ZP coefficients.
To show the accuracy of the semi-analytic beam models created from these coefficients, provided in {\tt EIDOS}, we reconstruct the models for 155 channels from 900 to 1670 MHz and calculate their NRMSE with respect to the given measurement and simulation.
These errors are also compared with the corresponding error of the PC-based models.

Fig. \ref{f:mse} shows the NRMSE of the PC and ZP-based models for the $E_{00}$ and $E_{01}$ elements.
The left and right panels show the errors for $\mathbf{E}^h$ and $\mathbf{E}^e$, respectively.
Although we have not described the modelling of the off-diagonal Jones elements, their NRMSE is included in this plot in order to compare them with the corresponding errors of the diagonal elements.
For example, we see that the NRMSE of $E_{01}^{ch}$ is almost the same as the NRMSE of $E_{00}^{zh}$ which shows that PC models are usually much closer to a given data at the expense of modelling the noise.
$E_{00}^{zh}$ is almost an order of magnitude higher than that of $E_{00}^{ch}$ which also reiterates the results presented in Fig. \ref{f:nrmse} for a single channel.
However, we need much less information to reconstruct beam models using ZP and, hence, only Zernike coefficients are provided in {\tt EIDOS}.
Errors of the diagonal elements increase with frequency for both $\mathbf{E}^h$ and $\mathbf{E}^e$ because there are more sidelobes at higher frequencies making the modelling less optimal.
Off-diagonal elements exhibit comparatively higher error, as expected.
The outlier points in the left panel are caused by RFI which is present in the AH measurement, but not in our models.

By comparing the left and right panels of Fig. \ref{f:mse}, one can see that PCs model the AH measurement and EM simulation with similar errors (blue and green markers), but ZPs exhibit higher error (red and magenta markers) in modelling the AH dataset.
This is because the measurements have low-level noisy structure and PCs are more prone to modelling the noise than ZPs.
Even though PCs are more faithful to a given data, representing all the structures and irregularities in an AH observation might not be desirable.

\subsubsection{Beamwidth and ripple} \label{s:ripple}
As a final test of the ZP-based models, we compare the full width at half-maximum (FWHM or `beamwidth') of the beam models with that of the original datasets as a function of frequency.
To calculate the beamwidth, we fit 2D elliptical Gaussian functions to the datasets and models within the region where beam amplitude is more than $0.01$.
The same method was used to calculate squint as described in Section \ref{s:ah} and shown in Fig. \ref{f:squint}.
The resulting beamwidth $\theta$ has a horizontal (semi-major axis) and a vertical (semi-minor axis) component and they vary from the theoretical beamwidth $\theta_t=\lambda/D$ (where $\lambda$ denotes wavelength and $D=14$ m, the dish diameter).
Fig. \ref{f:ripple} shows $\theta/\theta_t$ (for the horizontal width only) as a function of frequency for the Stokes I beams ($00$ element of the Mueller matrix) of the AH measurement (red dots), EM simulation (blue line), and the Zernike models created from the AH (red line) and EM (blue line) datasets.
The ripple of $\theta(\nu)$ is clearly visible here because $\theta_t(\nu)$ is smooth.
It is more prominent at the lower part of the band.
The models follow the original datasets closely, although they diverge more at the upper part of the band.
There is a clear division between the lower and upper parts of the L-band in terms of beamwidth and, hence, the band can be conveniently divided into two subbands: one before $\sim$ 1350 MHz, the other after it.
If we model the lower and upper subbands separately, the models will be more accurate, albeit at the expense of increasing the amount of required information.
For continuum science, a continuous full-bandwidth beam is more desirable, but for emission line studies, e. g. in case of HI intensity mapping, we can use a more accurate model of the beam created from the lower subband.

The ripple of the beamwidth is caused by the interference of the electric field diffracted from the secondary reflector with the electric field of the main beam \citep{deVilliers2013}.
To show the amplitude of the MeerKAT ripple, we subtract a smooth third-order polynomial from $\theta(\nu)$ calculated from the original AH and EM datasets.
The resulting ripple $\delta\theta(\nu)$ for the lower subband is shown in the inset of Fig. \ref{f:ripple} in arcmin units.
The top and bottom panels show the horizontal ($\delta\theta^H$) and vertical ($\delta\theta^V$) ripples, respectively, and the AH and EM datasets are represented by the red and blue colours, respectively.
$\delta\theta^H$ is significantly larger than $\delta\theta^V$ which can be interpreted as a frequency-dependent `beam squash' \citep{Heiles2001}, i. e. the beam is not squeezed symmetrically but squashed along a preferred direction.
The ripples predicted by the EM simulation match reasonably well with the AH measurements.
The match is relatively poor in case of $\delta\theta^V$ because it is at a lower level and, hence, more affected by the noise of the AH dataset.

\begin{figure*}
	\centering
	\includegraphics[width=0.45\linewidth]{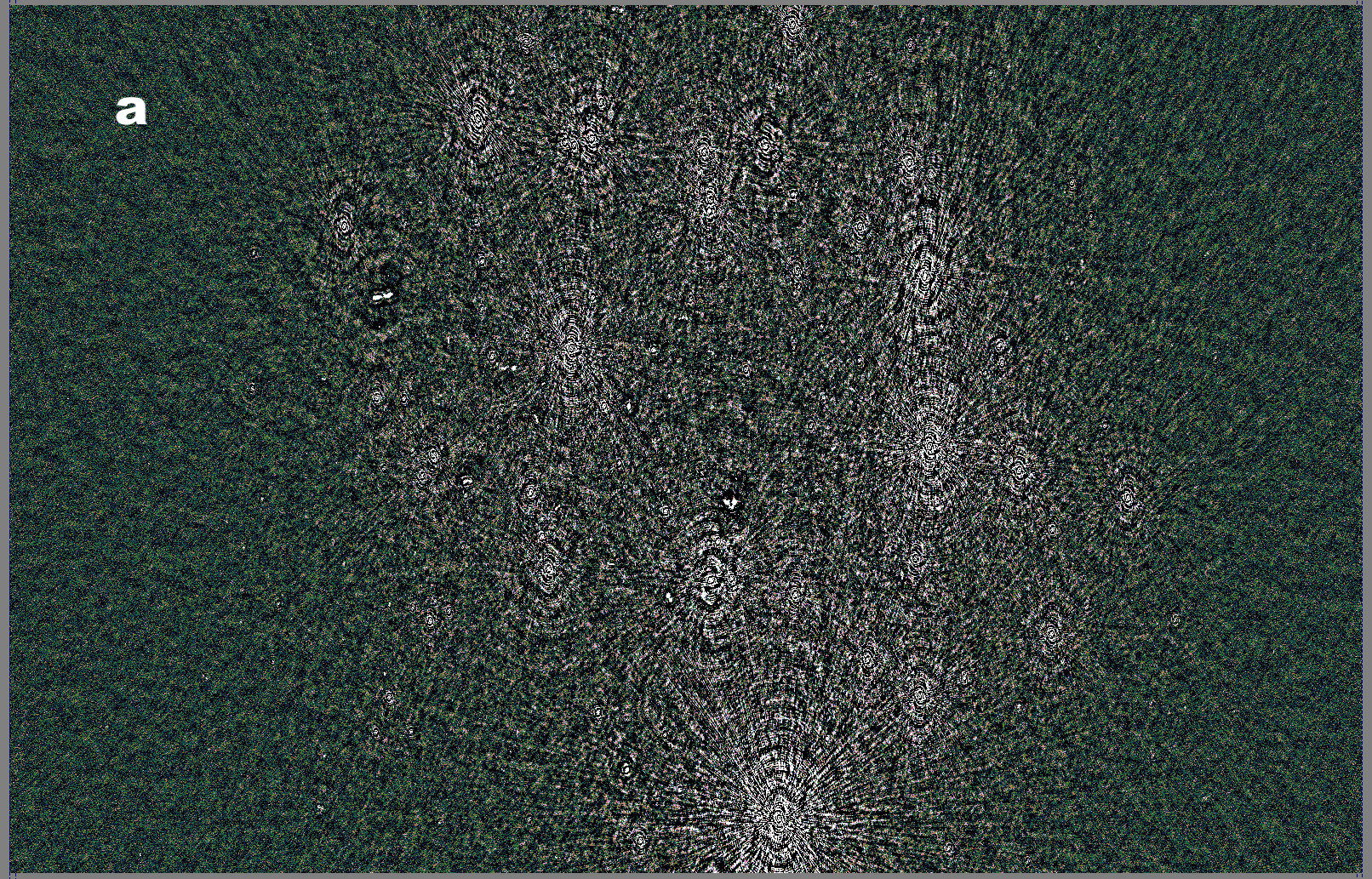}
	\includegraphics[width=0.45\linewidth]{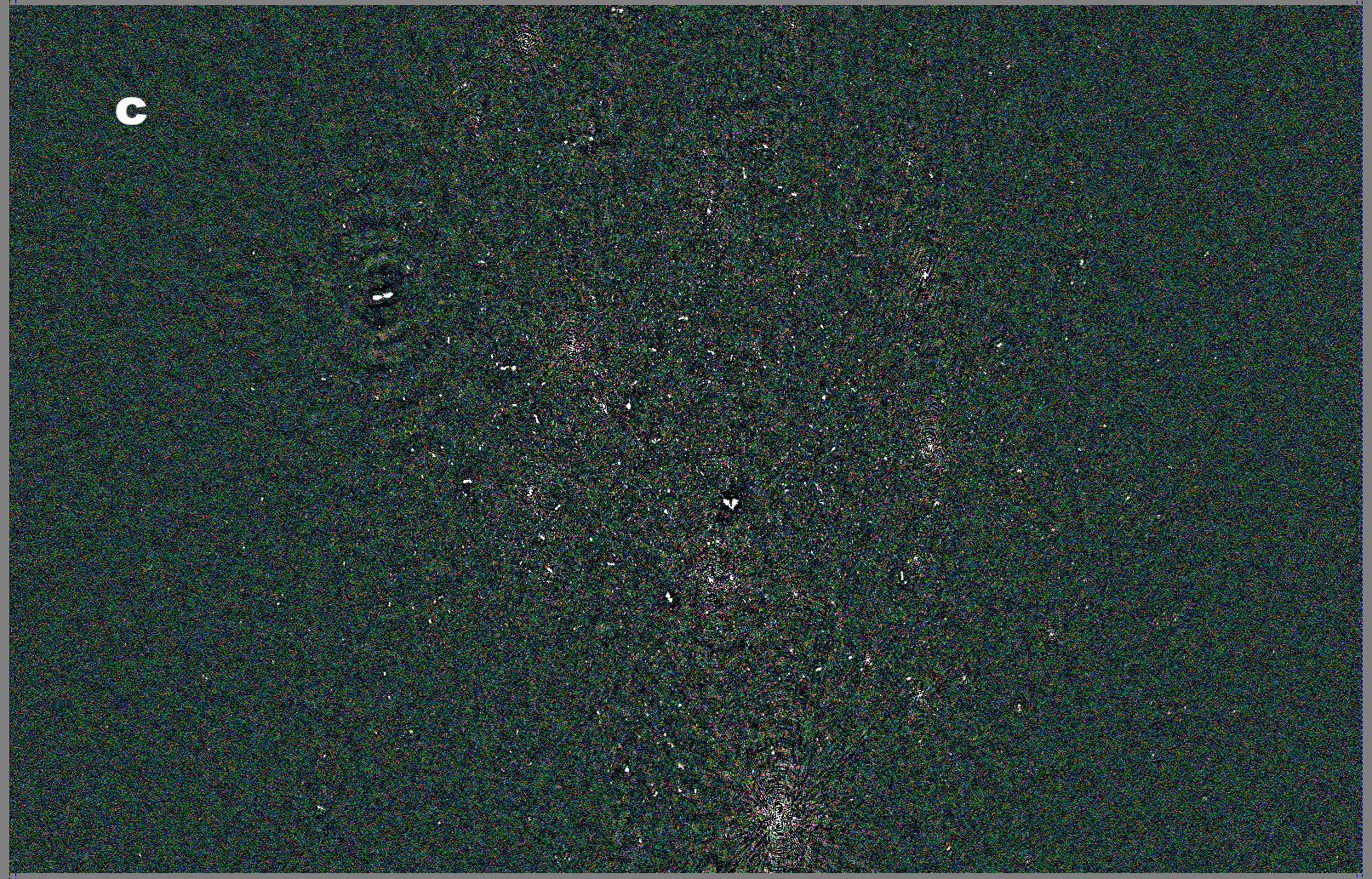}

	\includegraphics[width=0.45\linewidth]{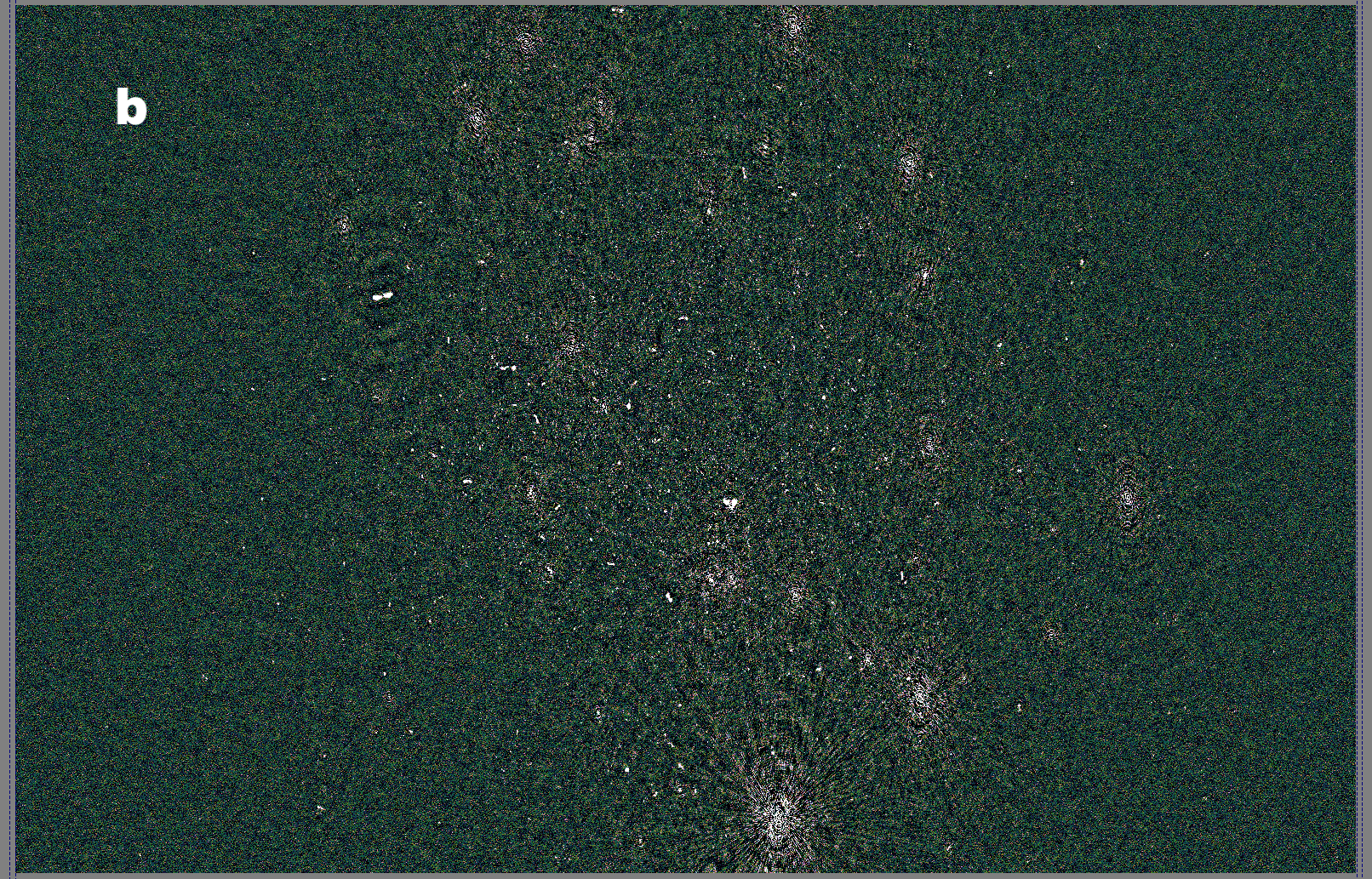}
	\includegraphics[width=0.45\linewidth]{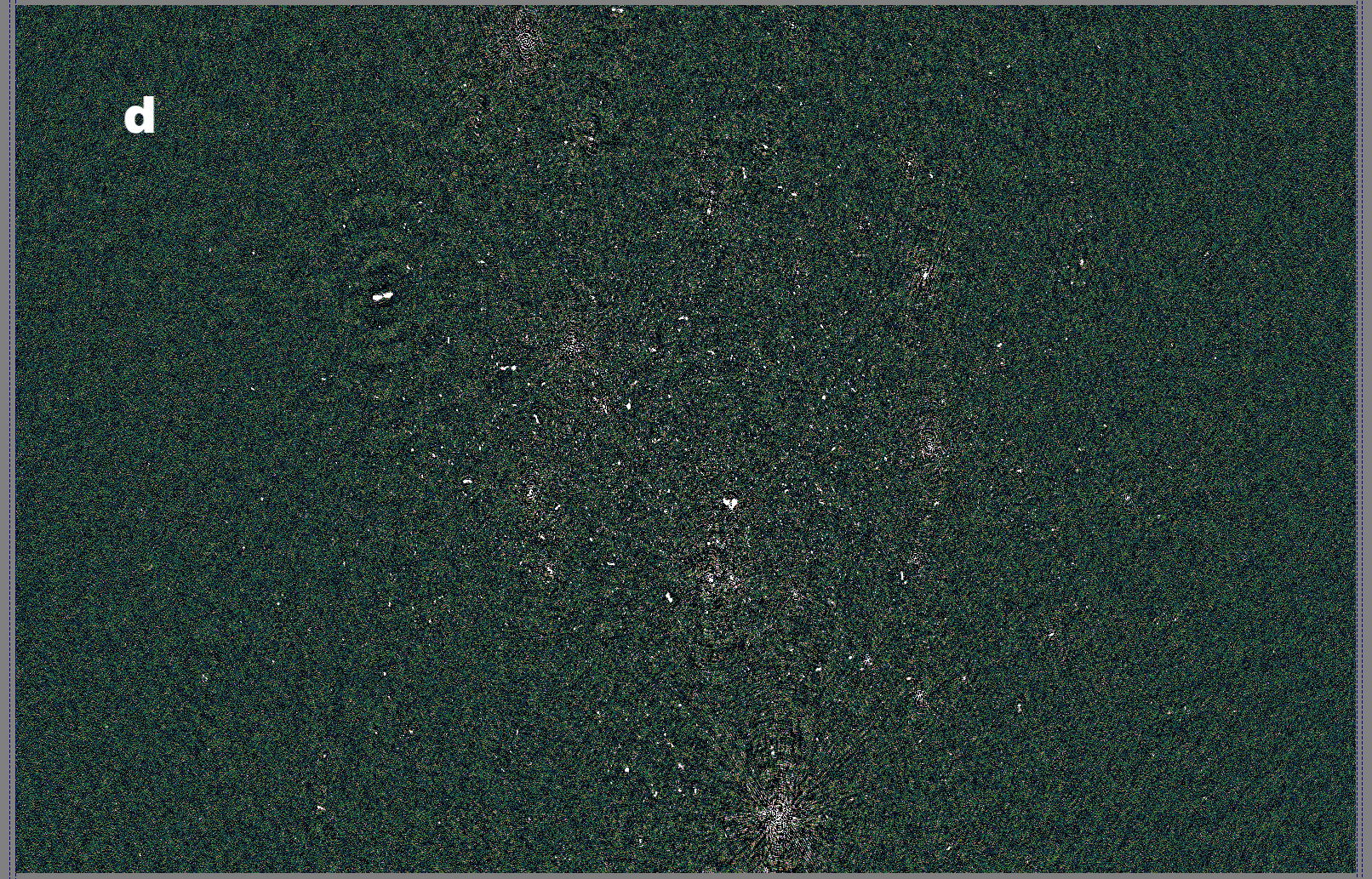}
	
	\caption{Images of the field around NGC4993, observed with 16 antennas of MeerKAT at L-band. The images are highly saturated (-10 $\mu$Jy to 50 $\mu$Jy linear scale) to emphasize deconvolution artefacts. Panel (a) has been produced using WSCLEAN, with a 1st-order polynomial spectral model, (b) using DDFacet without a primary beam, with a power-law spectral model, (c) using WSCLEAN, with a 3rd-order polynomial spectral model, and (d) using DDFacet with a primary beam, with a power-law spectral model.}
	\label{f:pbc}
\end{figure*}

\section{Primary beam correction}
We have provided the Zernike polynomials based model of the primary beam in the {\tt EIDOS} package.
The model is accurate for the diagonal elements of the Jones matrix.
It is averaged over three antennas.
In spite of these limitations, the model can already be used in primary beam correction of Stokes I images and in direction-dependent calibration.
As an example, we show an application of the {\tt EIDOS} primary beam using the {\tt DDFacet}\footnote{https://github.com/saopicc/DDFacet} \citep{Tasse2018} imaging software and compare the result with an image produced by {\tt WSCLEAN}\footnote{https://sourceforge.net/projects/wsclean/} \citep{Offringa2014} without any model of the primary beam.

Fig. \ref{f:pbc} demonstrates the application of the primary beam pattern during deconvolution of a wideband MeerKAT-16 (taken using 16 antennas of the MeerKAT array) image.
Panels (b) and (d) show images produced by {\tt DDFacet} with and without a primary beam model. In both cases, the underlying sky model being fitted during deconvolution is a power-law spectrum\footnote{This uses the SSD deconvolution mode of DDFacet} (i. e. two parameters, flux and spectral index, per model pixel). In case (b), this sky model is unable to account for the apparent spectral curvature induced by the primary beam, resulting in clear artefacts around brighter sources. In case (d), the imager is supplied with our MeerKAT primary beam model. The spectral effects of the beam are then accounted for properly, allowing the power-law spectrum sky model to fit the underlying emission better. The remaining artefacts are probably due to the unknown pointing errors.
Panels (a) and (c) show images produced by {\tt WSCLEAN} which deconvolves without any primary beam information. In case (a), the spectral model is a 1st-order polynomial (two parameters per model pixel), which also leaves substantial artefacts. In case (c), the spectral model is a 3rd-order polynomial (four parameters per model pixel), which is able to fully fit the spectral curvature induced by the beam.
Although the image quality in case (c) and (d) is comparable, the former uses twice as many degrees of freedom in the deconvolution model (being forced to absorb primary beam effects into the sky model), while the latter is able to recover a more physical sky model, with primary beam effects being explicitly accounted for in the instrumental measurement equation.

\section{Conclusion}
This is the second in a series of papers dealing with primary beam modelling effects of radio astronomy antennas.
The basic formalism was set in \textit{Paper I} \citep{Iheanetu2019} in the context of the VLA.
In this paper, we extend that formalism and use it to create MeerKAT beam models from astro-holographic (AH; $\mathbf{E}^h$) observations and EM simulations ($\mathbf{E}^e$).
Our aim was to present a pipeline that is able to create sparse representations of the beam, given any dataset, that can be used for direction dependent calibration and imaging.
This model can be called `eidetic' because it reconstructs the given datasets faithfully.
We have modelled the given $\mathbf{E}^h$ (Fig. \ref{f:ah}) and $\mathbf{E}^e$ (Fig. \ref{f:em}) over a diameter of 10 degrees using characteristic and analytic basis functions, for all frequencies of L-band and without considering the cross-polarization, and compared the results.
Here, $\mathbf{E}^h$ is averaged over three antennas and $\mathbf{E}^e$ is a generic simulation for an ideal antenna.

The diagonal elements of $\mathbf{E}^e$ matches well with $\mathbf{E}^h$ but their difference increases with frequency (Fig. \ref{f:em} and \ref{f:nrmse-em}).
We have decomposed the two beam datasets using principal components (PC), Zernike polynomials (ZP) and spherical harmonics (SH).
PCs and ZPs can represent the beam measurements and simulations very well with around 15 modes (compare the panels of Fig. \ref{f:Ec}), but 10-degree-beams cannot be modelled as well using the same number of SH modes (Fig. \ref{f:Es}).
Therefore, we focus on PCs and ZPs.

The coefficients corresponding to the most dominant PC and ZP modes are modelled in frequency using DCT.
Therefore, a full-bandwidth full-polarization 10-degree beam model can be represented using $2\times 2\times 2\times N^d\times N^b$ coefficients where $N^d$ is the number of DCT coefficients needed to model the spectral behaviour of each of the $N^b$ PC or ZP coefficients.
The first PC models the beamwidth as a function of frequency to a large extent (red line in the diagonal elements of Fig. \ref{f:s-pca}) and the small-scale ripple of the width can be seen clearly in the imaginary part of this component.
The spectral behaviour of the most dominant PC (Fig. \ref{f:s-pca}) and ZP (Fig. \ref{f:s-zern}) modes representing $\mathbf{E}^h$ (dots) and $\mathbf{E}^e$ (solid lines) matches well in the diagonal elements.

Beam models can be represented by analytic basis functions like ZPs using less information than the characteristic bases like PCs because for the latter, we also need to store the eigenbeams.
By looking at the 15 most dominant ZP modes (Fig. \ref{f:zern}), we see that the diagonal elements of $\mathbf{E}^h$ or $\mathbf{E}^e$ are modelled mainly by the symmetric spherical modes and the asymmetries in the beam are modelled by the coma and astigmatism modes.

We have been able to represent the spectral behaviour of the coefficients using 40 DCT coefficients for $\mathbf{E}^h$ and 30 for $\mathbf{E}^e$ (Fig. \ref{f:dct}).
Before DCT decomposition, the coefficients for the RFI-affected channels in $\mathbf{E}^h$ are calculated by interpolating from the RFI-free channels.
Through interpolation and DCT, we have calculated coefficients for all frequencies between 900 and 1670 MHz with a resolution of 0.1 MHz.

The resulting beam models calculated from the ZP coefficients have a residual error (after subtracting the model from data) of around $10^{-5}$ corresponding to the \textit{power} of the diagonal element (Fig. \ref{f:Ec}).
For the electric field, the corresponding residual levels would be around $10^{-3}$.
Comparatively, the PC-based models exhibit lower residuals (Fig. \ref{f:Ec}).
The off-diagonal elements are less accurate and the errors of all the Jones elements increase with frequency, as seen in the trend of the normalised root-mean-square error (NRMSE) (Fig. \ref{f:nrmse} and \ref{f:mse}).
However, NRMSE should be used with caution because it is averaged over a diameter of 10 degrees and, hence, mostly dominated by errors in the nulls and noisy regions.

The antenna and frequency dependent squints of the beam (Fig. \ref{f:squint}) were taken out before modelling, but they can be reintroduced by re-centring the model beams at a later stage if required.
On the other hand, the squash effect (Fig. \ref{f:ripple}) is already incorporated in the Zernike coefficients; the beamwidth has a ripple of amplitude $\sim 0.3$ arcmin in the horizontal and $\sim 0.1$ arcmin in the vertical direction and it varies as a function of frequency with a period of $\sim 20$ MHz.

The ZP and corresponding DCT coefficients for representing $\mathbf{E}^h$ and $\mathbf{E}^e$ are provided in the {\tt EIDOS} software tool that can be used to calculate MeerKAT eidetic beam models for any frequency of the L-band.
The diagonal Jones terms have been modelled with a high degree of certainty; between the high signal-to-noise ratio of the AH measurements and the excellent match to EM, we do not expect that the diagonal model can be much improved (although further investigation into elevation dependence and antenna-to-antenna differences is certainly merited).
The off-diagonal terms, however, need further refinement; the intrinsically high noise of AH suggests that these models can be improved with additional AH observations, and relatively poor match with EM simulation also needs to be further investigated.
Therefore, we have not discussed the modelling of the off-diagonal elements in this paper.

The advantage of having a Zernike-based eidetic model of the beam is that it can be calculated quickly and instantaneously during calibration or beam-correction.
For example, we have compared the quality of the images produced by applying our eidetic beam on a real observation using the {\tt DDFacet} imager with that produced by {\tt WSCLEAN} without any prior information of the beam.
We have seen that {\tt WSCLEAN} can produce images comparable in quality to that of {\tt DDFacet} (Fig. \ref{f:pbc}), but only at the expense of twice as many degrees of freedom.
Therefore, in spite of the limitations, the beam models given in {\tt EIDOS} can already be used in primary beam correction of Stokes I images through {\tt DDFacet}.

Our beam models can be used in MeerKAT-specific pipelines through the radio astronomy package {\tt STIMELA}.\footnote{\url{https://github.com/SpheMakh/Stimela}}
The VLA beam models presented in Paper I will also be included in this tool.
In future papers, we will explore the spatial, spectral and polarization effects of these beams on continuum imaging and intensity mapping.

\section*{Acknowledgements}
The MeerKAT telescope is operated by the South African Radio Astronomy Observatory (SARAO), which is a facility of the National Research Foundation (NRF), an agency of the Department of Science and Technology (DST) of South Africa.
KMBA and MGS acknowledge support from SARAO and NRF (grant 84156) and DILV from NRF grant 75322.
The research of OS and KI is supported by the South African Research Chairs Initiative of the DST and NRF.
We thank Christopher Jackson Finlay (SARAO) for the parallelization script in {\tt EIDOS}.
We also thank Kavilan Moodley for useful suggestions.

\section*{Data availability}
The beam models presented in this paper can be created using the EIDOS python package: \url{https://github.com/ratt-ru/eidos}. The underlying data is available in the SARAO Archive: \url{https://archive.sarao.ac.za}.


\bibliographystyle{mnras}
\bibliography{ref}


\appendix
\section{Zernike polynomials}  \label{ap:ZP}

The Zernike polynomials of order $n$ and angular frequency $m$ can be written in polar coordinates as \citep[following][chapter IX, section 9.2.1]{Born1999}
\begin{equation}
Z_n^m(\rho,\phi) = R_n^m(\rho) e^{im\phi}
\end{equation}
where $n$ and $m$ are non-negative integers with $n\ge |m|$ and $n-|m|$ is always even. The radial polynomials
\begin{equation}
R_n^{\pm m}(\rho) = \sum\limits_{s=0}^{\frac{n-m}{2}} \frac{(-1)^s (n-s)!}{s!\left(\frac{n+m}{2}-s\right)! \left(\frac{n-m}{2}-s\right)!} \rho^{n-2s}
\end{equation}
and the normalisation is such that for all permissible values of $n$ and $m$, $R_n^{\pm m}(1)=1$.

\bsp
\label{lastpage}
\end{document}